\documentclass[aps,pra]{revtex4}

\usepackage{amssymb}
\usepackage{mathtools}
\usepackage{physics}

\begin{document}
\title{Depictions of Quantum Reality in Kent's Interpretation of Quantum Theory}

\author{Brendan \surname{Marsh}}
\email{marshbp@stanford.edu} 
\altaffiliation[Current address: ]{Department of Applied Physics and Ginzton Laboratory, Stanford University, Stanford, CA 94305}
\affiliation{Department of Applied Mathematics and Theoretical Physics, University of Cambridge, Wilberforce Road, Cambridge, CB3 0WA, U.K.}

\begin{abstract}

At present, quantum theory leaves unsettled which quantities ontologically, physically exist in a quantum system. Do observables such as energy and position have meaningful values only at the precise moment of measurement, as in the Copenhagen interpretation? Or is position always definite and guided by the wave function, as in de Broglie-Bohm pilot wave theory? In the language of Bell, what are the ``beables" of quantum theory and what values may they take in space and time? This is the quantum reality problem. A definitive answer requires not just describing which physical quantities exist in a quantum system, but describing what configurations of those quantities in space and time are allowed, and with what probability those configurations occur. Adrian Kent sets out a new vision of quantum theory along these lines. His interpretation supplements quantum theory to infer the value of physical quantities in spacetime from the asymptotic late-time behavior of the quantum system. In doing so, a Lorentz-covariant and single-world solution to the quantum reality problem is achieved. In this paper, the framework of Kent's interpretation is presented from the ground up. After a broad overview, a derivation of the generalized Aharonov-Bergmann-Lebowitz (ABL) rule is provided before applying Kent's interpretation to toy model systems, in both relativistic and non-relativistic settings. By adding figures and discussion, a broad introduction is provided to Kent's proposed interpretation of quantum theory. 

\end{abstract}

\maketitle

\vspace{-6mm}
\tableofcontents
\normalsize

\newpage
\section{Problems with Measurement and Quantum Reality}
\label{Introduction}

Quantum theory was a resounding success of twentieth century physics. The quantum revolution offered far-reaching insights into physical reality, finding relevance not just in the minds of physicists but of mathematicians, philosophers, chemists, even engineers and computer scientists. Much is understood, especially about empirical aspects of quantum mechanics, yet much remains mysterious. Our empirical knowledge of quantum mechanics is, by definition, limited to what can be observed, and many interesting questions lie beyond current experimental reach. In particular, interpretations of quantum theory, the role of measurement itself, and ``peaceful coexistence" between quantum theory and Einstein's theory of relativity, all pose ongoing questions with deep physical significance.

The interpretation of quantum theory is an especially rich subject which plays host to many rival schools of thought. Motivating the need for these competing interpretations are problems encountered with the once-standard Copenhagen interpretation. Although pragmatic and highly effective in the laboratory, the Copenhagen interpretation posits a perhaps unnatural mixture of deterministic evolution of quantum states and non-deterministic measurement events, a feature seen by many to hint at the existence of a deeper interpretation. Very close to the problems of the Copenhagen interpretation is the quantum measurement problem, as well as the more modern quantum reality problem, as described by Kent \cite{RealWorld, MainKent}. Let us formulate these problems and expose some of the holes in the Copenhagen interpretation.

\subsection{The Quantum Measurement Problem}

Two of the fundamental postulates of Copenhagen quantum theory can come in conflict, with their coexistence dependent on imprecise boundaries between the quantum system, the observer, and their interaction via measurement. Attempting to unify observers and quantum systems under a single quantum umbrella quickly leads to trouble. Consider the following two postulates of Copenhagen quantum theory.

{\centering
\vspace{5mm}
\textit{Postulate: The time evolution of quantum states is governed by the Schr\"odinger equation.}
\vspace{5mm} \par
}

In the Schr\"odinger picture of quantum mechanics, a closed quantum system is described by a ket vector $\ket{\Psi(t)}$ which fully describes the state of the system and which evolves in time. It is a fundamental postulate that the time evolution of this quantum state is described by the Schr\"odinger equation,
\begin{equation}
i\hbar\frac{\partial}{\partial t}\ket{\Psi(t)} = H\ket{\Psi(t)},
\end{equation}
where $H$ is the Hamiltonian operator, describing the energy of the quantum system. In particular, for Hamiltonians which do not explicitly depend on time, the time-evolved state of the quantum system can formally be written as
\begin{equation}
\ket{\Psi(t)} = e^{-itH/\hbar}\ket{\Psi(0)},
\end{equation}
where the exponentiated operator $e^{-itH/\hbar}$ is defined in terms of its Taylor series. This operator is \textit{unitary}, so the time evolved state $\ket{\Psi(t)}$ for any time $t$ is related to the initial state $\ket{\Psi(0)}$ by some unitary transformation. In fact, unitary time evolution holds for systems with time-dependent Hamiltonians as well. Thus, the Schr\"odinger equation implies that all closed quantum systems evolve unitarily in time.

In many ways, unitary time evolution is a very well-behaved way in which a system can change over time. First, it is deterministic: given the initial state $\ket{\Psi(0)}$ and knowledge of the Hamiltonian of the system, one can prescribe with certainty what the state of the system will be at any future time --- there is no randomness here. Second, unitary time evolution is reversible in the sense that for any unitary operator $U(t)$ there exists an inverse operator given by the conjugate transpose, so that $U(t)U(t)^\dagger=U(t)U(t)^{-1}=I$. In particular, this implies that given the state of the system $\ket{\Psi(t')}$ at some time $t'$, there exists an inverse operator $U(t')^{-1}$ which allows one to recover the initial state: $\ket{\Psi(0)}=U(t')^{-1}\ket{\Psi(t')}$.

{\centering
\vspace{5mm}
\textit{Postulate: The outcome of measuring a quantum state is random.}
\vspace{5mm} \par
}

Measurement of a quantum state, on the other hand, is quite a different beast. For an outside observer to \textit{learn} something about a quantum state requires the observer to intrude on the quantum system, which we previously considered as a closed system, and probe some part of it. The intrusion of the observer significantly changes the dynamics of the quantum system.

In their simplest form, measurements are represented by Hermitian operators $M=M^\dagger$ acting on the Hilbert space of the quantum state. By the spectral theorem, the eigenvalues $\{m_i\}$ of $M$ are all real numbers and are postulated to correspond to the possible outcomes of performing the measurement. With any eigenvalue $m_i$ we can associate a projection operator $P_i$ which projects the quantum state onto the eigenspace of $m_i$.

At the point of the measurement, it is postulated that the unitary time evolution prescribed by the Schr\"odinger equation ceases, temporarily being replaced by a single randomly chosen projection defined by an operator $P_i$. The projection operator is chosen at random, corresponding to a randomly chosen outcome of the measurement, but chosen with well-defined probabilities. The probability of obtaining the measurement outcome $m_i$ is prescribed by the Born rule:
\begin{equation}
\textrm{Pr}\left( m_i \right) = \bra{\Psi}P_i\ket{\Psi},
\end{equation}
where $\ket{\Psi}$ is the quantum state just prior measurement. Supposing upon measurement the outcome was found to be $m_i$, the post-measurement quantum state $\ket{\Psi'}$ is then found to be the pre-measurement state $\ket{\Psi}$ projected onto the eigenspace of $m_i$,
\begin{equation}
\ket{\Psi}\to \ket{\Psi'}= N P_i\ket{\Psi}
\end{equation}
where $N$ is a normalization factor to ensure that the post-measurement state is normalized so that $\bra{\Psi'}\ket{\Psi'}=1$.

Time evolution given by projection transformations, as postulated for measurement processes, is \textit{very} different from the time evolution by unitary transformations seen for the closed quantum system. Projections are non-unitary transformations, they irrevocably discard any component of the quantum state which is orthogonal to the projection eigenspace. Once the projection occurs, there is no going back --- in general there exists no inverse operator for projections. Thus, by postulate, it would seem that measurement in quantum mechanics is both non-deterministic, as the projection operator is chosen at random, and irreversible.

{\centering
\vspace{5mm}
\textit{Are these postulates consistent?}
\vspace{5mm} \par
}

There is an awkward tension here between closed quantum systems, which behave nicely under Schr\"odinger equation unitary evolution, and observers trying to peak in at the quantum system and learn about them, apparently inducing non-unitary projections. It seems that the type of time evolution that a quantum system undergoes depends critically on whether the system is being ``measured" by an outside observer.

While observers may take the form of macroscopic systems, they are nonetheless a macroscopic collection of interacting microscopic systems whose dynamics ought to be subject to quantum mechanics. But isn't the total system containing the observer \textit{and} the quantum system being measured just another, larger, closed quantum system? In this situation, the first postulate mandates that the total quantum system, including the subsystem being measured, evolve unitarily. On the other hand, a measurement is taking place on the subsystem by the observer, so the second postulate mandates that the subsystem should undergo a random projection. Well, which one is it? The quantum state being measured cannot simultaneous undergo unitary evolution and projective evolution: the two transformations are mathematically and physically incompatible with one another!

Have we reached a contradiction in the fundamental postulates of Copenhagen quantum mechanics? This gedankenexperiment seems to imply that either macroscopic observers are \textit{not} quantum systems or that measurements are in fact unitary processes. Such is the measurement problem. While the Copenhagen interpretation accepts as physical law both the Born rule and the projection postulate for measurement, other interpretations do not --- avoiding the ensuing catastrophic clash with unitary Schr\"odinger evolution.

\subsection{The Quantum Reality Problem}

The quantum measurement problem addresses what happens in quantum systems during measurements to make them appear stochastic, when the Schr\"odinger equation tells us they \textit{should} behave deterministically. The quantum reality problem goes further, addressing the value of physical quantities in a quantum system not just during measurement events but at all points in space and time. The problem is firstly that quantum theory leaves open which quantities in a quantum system are granted ``physical" status to begin with. In the language of John Bell's ``beables" \cite{Beables,Bell}, emphasizing what can actually \textit{be} in a quantum system rather than what can merely be \textit{observed} through measurement, there does not exist any final word on a precise description of the beables of quantum theory. Moreover, given a particular closed quantum system, there is no precise description of how those beables may be distributed in space and time.

A successful solution to the quantum reality problem should imply the highly verified predictions of Copenhagen quantum theory in the small scale domain of quantum mechanics, including the emergence of the Born rule in quantum measurements, as well as the quasiclassical physics of macroscopic objects, which appear to follow mostly deterministic equations of motion. If that all sounds too easy (!), then one may additionally seek a solution to the \textit{Lorentzian} quantum reality problem --- that is, a solution to the quantum reality problem which provides a Lorentz covariant description of beables and physical reality, respecting special relativity.

A full solution to the quantum reality problem consists of specifying: 

\vspace{2mm}

 1. A set of physical quantities, a set of beables, which describe the physical reality of a quantum system. 

\vspace{2mm}

 2. A sample space of all possible configurations of those beables in spacetime, what one might call histories. Or, at least, a sample space of mathematical objects from which the beable configuration can be inferred. 

\vspace{2mm}

 3. A probability distribution over the sample space. 

\vspace{2mm}

Then, the \textit{actual} physical reality of a quantum system can be postulated to correspond to a single possible history of beables, thus a single configuration of physical quantities in space and time, chosen at random from the distribution.

It is worth noting that finding \textit{a} solution to the quantum reality problem does not imply finding \textit{the} solution to the quantum reality problem. That is, a specification of a set of beables and a coherent description of their possible values in spacetime indeed defines a physical reality of the quantum system, but that description of physical reality need not be the fundamentally correct one. It may ultimately be found that a specific set of beables and a sample space of configurations of those beables, i.e. a specific solution to the reality problem, is the one realized in Nature. For now, it is challenging enough to construct \textit{any} solutions to the reality problem which sketch a coherent and Lorentz covariant picture of physical reality, and reproduce both the empirical results of quantum mechanics as well as those of quasiclassical mechanics in their respective domains.

Different interpretations of quantum theory indeed address quantum reality differently. The pilot wave theory of de Broglie and Bohm \cite{Bohm} provides a solution to the quantum reality problem, suggesting a pilot wave which steers a real particle with a definite position in accordance with a guidance equation. Thus, the pilot wave theory suggests particle position as a fundamental beable, with well-defined histories, though for relativistic systems, in which particle number may not be conserved, no thoroughly successful extension has been found. On the other hand, the many-worlds interpretations stemming from the work of Hugh Everett \cite{Everett,KentAgainstManyWorlds} describe a fantastically different kind of physical reality with a constantly branching universal wave function. Even though the many-worlds interpretation may indeed be a valid interpretation, it remains of interest whether it is a necessary construct for describing quantum reality. That is, does there exist a single-world, Lorentz-covariant solution to the quantum reality problem?

In recent years, Adrian Kent has given us a start on such a solution to the Lorentzian, single-world quantum reality problem \cite{MainKent,Kent20140241,PhysRevA.96.062121,RealWorld}. In his interpretation, it is assumed from the start that there indeed exists a more fundamental relativistic quantum theory which rigorously describes the physics of measurement and interaction in a unitary fashion, despite such a theory not being known at present. In Kent's proposed interpretation, one supplements such a quantum theory with beables in spacetime, the configurations of which are inferred from an asymptotic late time distribution. To every late time beable distribution there is an associated probability, corresponding to the probability of measuring such a distribution given the unitarily evolved final universal quantum state. The beables themselves are often stress-energy distributions, though a number of alternatives have been considered. Taken together with a few important assumptions on the late-time behavior of the quantum system, Kent has shown that this form of solution holds promise as an extension of relativistic quantum theory which solves the Lorentzian quantum reality problem.

In this paper, Kent's framework for solving the quantum reality problem is first presented in words and figures for both non-relativistic and relativistic systems, noting some major differences between the two. In section \ref{framework}, the basic building blocks of the interpretation are established. Section \ref{ABL} follows with a derivation and generalization of the Aharonov-Bergmann-Lebowitz rule, as will be needed for Kent's framework. Finally, sections \ref{NonRel} and \ref{RelSolns} apply Kent's interpretation to various models for non-relativistic and relativistic systems, respectively. Adding to the presentation of the models are figures, discussion of interesting features in Kent's interpretation, and a new concept of a ``region of indeterminacy" which aids in building visual intuition for Kent's framework.  

\newpage
\section{The Framework of Kent's Interpretation}
\label{framework}

\begin{figure}[h]
    \center
    \includegraphics[]{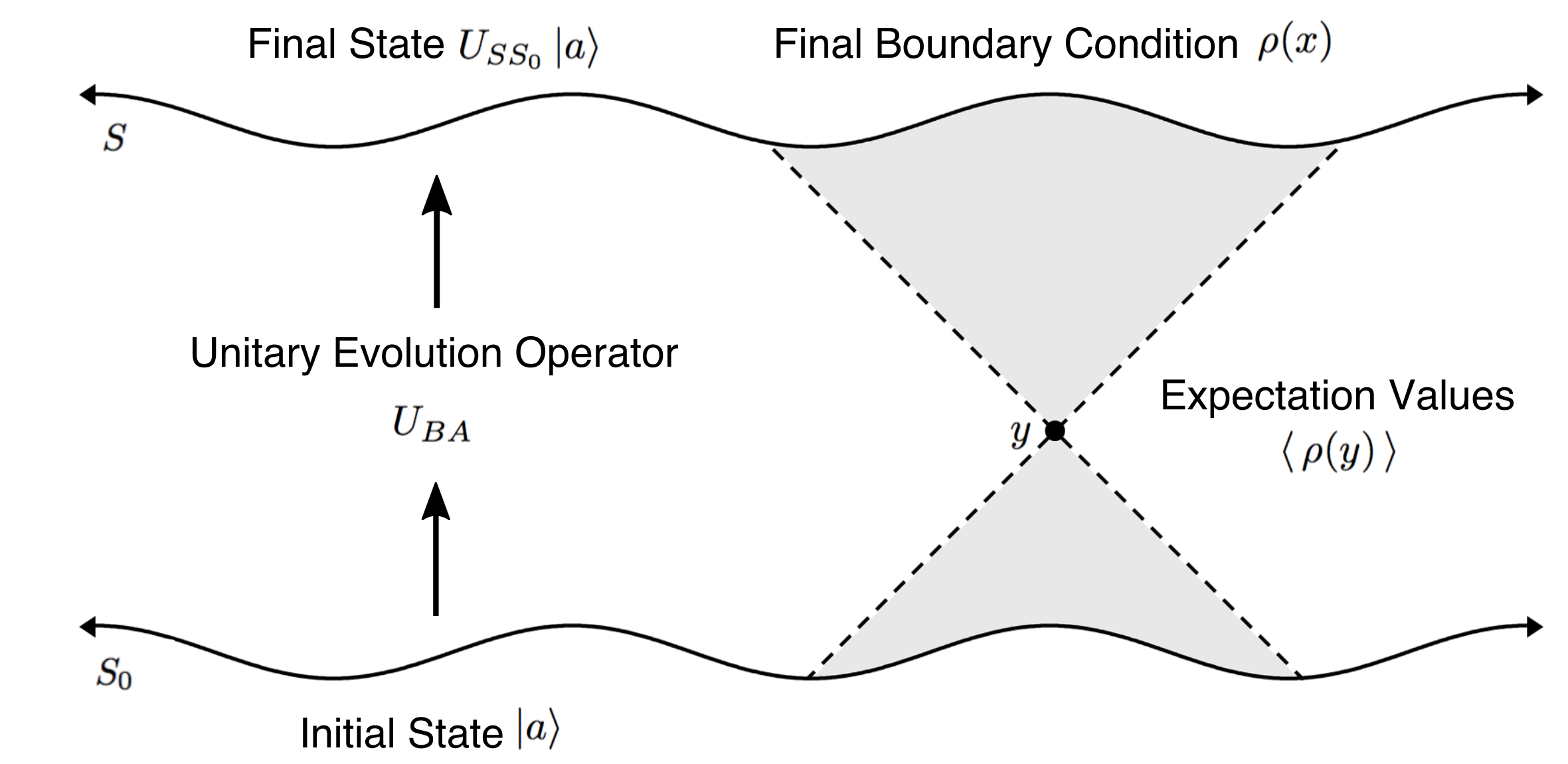}
    \caption{A depiction of Kent's framework for defining beable expectation values for any spacetime point $y$ located between an initial spacelike hypersurface $S_0$ and final spacelike hypersurface $S$. Shaded regions represent spacetime points within the future or past light cone from $y$, which are treated differently in relativistic solutions.}
    \label{OverviewFig}
\end{figure}

Before concerning ourselves with mathematical machinery, let us take a broad and visual approach to Kent's interpretation. To solve the quantum reality problem, one must specify an appropriate beable corresponding to a physical quantity, lay out the sample space of possible configurations of that beable in spacetime for a given quantum system, and assign probabilities to those configurations of beables. The framework can be split into four parts, A through D, which will be introduced in a natural order. These steps will be accompanied by figure \ref{OverviewFig}, painting an overview. Where the framework differs for relativistic and non-relativistic systems, each will be explained separately.

\iffalse
$$\ket{a}$$
$$U_{BA}$$
$$\ket{c}=U_{SS_0}\ket{a}$$
$$\rho (x)$$
$$\langle \,\rho (y)\, \rangle$$
\fi

\subsection{The Quantum System and its Initial Conditions} 
\label{KentStep1}
In non-relativistic models, space takes the form of three-dimensional Euclidean space, where the physics of the system is considered to begin at some initial time up and run until the asymptotic infinite future. In relativistic models the setting is Minkowski space, where all spacetime points between an initial spacelike hypersurface $S_0$ and a late-time spacelike hypersurface $S$ are considered, where $S$ is taken to asymptotically approach the infinite future. To be precise, $S$ being in the future of $S_0$ means that for all points $x\in S$ every inextendible past-directed, timelike curve through $x$ intersects $S_0$.

Kent's interpretation does not start from a blank slate. A quantum theory is presupposed which prescribes pure unitary evolution of the quantum state, including interactions and measurements without any break in that unitarity. As admitted by Kent throughout his work, to date there exists no such rigorous relativistic quantum theory. This makes a completely rigorous test of any proposed interpretation challenging, but it does not invalidate the foundational effort to resolve the quantum reality problem. Predicated on the existence of such a theory, Kent shows that one may \textit{supplement} the theory to give a mathematically precise and Lorentz covariant description of physical reality in a quantum system, solving the reality problem.

To complete the initial setup, it is supposed that the initial state of the system, $\ket{a}$ in figure \ref{OverviewFig}, is given on the initial surface: either the initial time slice $t=0$ of three-dimensional Euclidean space or at all points on the initial hypersurface $S_0$. 

\subsection{Unitary Evolution to a Final State} 
\label{KentStep2}
As a unitary law of evolution for quantum states of our system is provided by the presupposed quantum theory, the initial state $\ket{a}$ may be evolved in time. In the non-relativistic models, evolution is generated by the Hamiltonian $H$ for the system, so that states unitarily evolve via the evolution operator $e^{-iHt/\hbar}$. Evaluating for the final time $T$, which is taken in the limit as $T\to \infty$, provides the final state $e^{-iHT/\hbar}\ket{a}$ of the system. For relativistic scenarios, a unitary operator $U_{BA}$ for the evolution of quantum states between any two spacelike hypersurface A and B can be provided by the Tomonaga-Schwinger formalism \cite{TomonagaSchwinger}. Then, the initial state $\ket{a}$ can evolve according to a unitary operator $U_{SS_0}$, so that the quantum state on the late-time hypersurface $S$ is given by $U_{SS_0}\ket{a}$.

\subsection{A Final Boundary Condition} 
\label{KentStep3}
Thus far, Kent's framework is more or less ``quantum theory as usual". Now come the novel aspects which supplement the quantum theory, and which allow one to solve the quantum reality problem. Recalling the definition of the quantum reality problem, a sample space of possible histories of beables in spacetime needs to be defined, corresponding to possible physical realities, along with a probability distribution on the sample space. 

The quantum sample space of Kent's framework is composed of beable distributions on the late time hypersurface, referred to as \textit{asymptotic final boundary conditions} for the beables of the theory. That is, for non-relativistic systems, the beable configuration (such as a mass distribution) is specified for the system at the final time $T$, which is taken to approach the infinite future. In relativistic solutions, the final boundary condition specifies the beable configuration (think of the mass-energy distribution across space) on the final hypersurface $S$, taken in the limit as $S$ tends to the infinite future. In figure \ref{OverviewFig}, the final boundary configuration is denoted by $\rho(x)$ for all $x\in S$. Kent's claim is then that the specification of the final beable configuration is in fact sufficient to infer the expectation value of the beable configurations at all other points in spacetime: this challenge is addressed shortly!

The probability distribution for the final boundary conditions is naturally defined by standard Born rule probabilities. Given the unitarily evolved final state on the late-time hypersurface, one may consider a hypothetical measurement of the beable at all points on the late-time hypersurface (all these measurements commute since they are spacelike). The probability for measuring an outcome at any one point is governed by the Born rule. Thus, the probability density for measuring any final boundary condition may be provided by Born rule probabilities. One such final boundary condition is then chosen at random from the sample space of all possible final boundary conditions, weighted by its Born rule probability.

Mathematically, the possible final boundary conditions may be characterized as the possible outcomes of a simultaneous measurement of beables at all points $x\in S$ on the final hypersurface $S$. However, this is just a mathematical characterization, no physical measurement takes place - the quantum system is treated as a closed system, with no external observer present to make such a measurement. Such a measurement is therefore referred to as a \textit{fictitious late-time measurement}. As explained by Butterfield \cite{Butterfield}, one must think of this fictitious measurement in a ``tenseless" sense: the final boundary condition is not selected only upon reaching the late-time hypersurface, but is an inherent part of the quantum system, pre-selected at random to describe the final beable configuration. This means that there is no retroactive causation: the late-time boundary condition is not chosen at a final time, proceeding to influence the beables at earlier spacetime points. On the contrary, the late-time boundary condition in fact dictates the beable values, and thus the physical reality of the system, throughout all spacetime between $S_0$ and $S$.

A central postulate of Kent's framework is that there indeed exists a convergent asymptotic probability distribution of final boundary conditions, as time tends to future infinity. To be precise, Kent proposes not just one final hypersurface $S$, but a sequence of hypersurfaces $S_i$ which approach the infinite future as $i$ tends to infinity. That is, each $S_i$ is a Cauchy surface, analogous to an instant of time, and the sequence $S_i$ approaches the infinite future. For each $S_i$, one may consider a fictitious measurement of the beable at all points on $S_i$, and thus the probability of obtaining any given final beable distribution on $S_i$. The sequence of hypersurfaces $S_i$ thus define a sequence of beable probability distributions. Kent's proposal is that such a sequence defines a \textit{convergent sequence of probability distributions} as the $S_i$ tend to future infinity, so that the limiting probability distribution of beable configurations on those $S_i$ defines the sample space of possible final boundary conditions. A selection of a single final beable configuration from the sample space, using standard Born rule probabilities, will then define the expectation value of beables at all previous spacetime points, and thus define the physical reality of the system, as will be explained below. Keeping these points about convergent sequences of hypersurfaces in mind, for simplicity a single hypersurface $S$ is often considered which is taken in the limit of tending to the infinite future.

An important condition is required for the existence of a well-defined asymptotic final boundary condition. Although there exists non-trivial interactions between particles in the quantum system, \textit{the system must evolve so that particle interactions eventually become negligible} as time tends towards the infinite future. Otherwise, the system would not asymptotically approach any single state, and thus could not asymptotically approach a single probability distribution for final beable configurations. It is quite a non-trivial assumption that particle interactions in the universe eventually become negligible, and while some cosmological models suggest a future which is incompatible with this notion (big crunch scenarios: particle interactions become ever stronger), other current models \textit{are} compatible (e.g. big rip scenarios: particle interactions become ever weaker).

\subsection{Beable Expectation Values} 
\label{KentStep4}

The sample space of Kent's framework is composed of final boundary conditions for the beables of the quantum theory. How do these final boundary conditions imply the expectation values of beables at all other points in spacetime? For the non-relativistic solutions, the exact link between the initial state, final boundary condition, and expectation values of intermediate time beables is given by the Aharonov-Bergmann-Lebowitz (ABL) rule \cite{ABLpaper}. Since this link is central to Kent's theory, section \ref{ABL} is dedicated to properly deriving this result, but it is discussed informally here. The relativistic solutions formally do not make use of the ABL rule, for reasons to be made clear in section \ref{RelSolns}. However, the form of the expectation values will be very similar.

In short, the ABL rule prescribes the probability of an outcome of a given measurement B, at any given spacetime point $y$, given the initial state of the quantum system and given that the system later yields a given outcome of another measurement C. In the language of ABL, one arrives at such a probability by ``pre-selecting" the initial condition and ``post-selecting" for some final outcome. The applicability to Kent's solutions is then immediately obvious, given our initial state $\ket{a}$ and final boundary condition $\rho(x)$. By averaging over all possible beable values at the point $y$ weighted by the corresponding ABL probability, we finally construct the expectation value of the beable $\langle \,\rho (y)\, \rangle$ for every intermediate point in our spacetime. 

There is a significant difference in deriving beable expectation values for relativistic systems. In the non-relativistic solutions, Kent conditions on the probability of finite time beables by post-selecting on the entire final boundary condition for all spatial points at time $T$. However, in the relativistic solutions, in constructing expectation values at an intermediate spacetime point $y$, Kent post-selects not on the final boundary condition for the beables at all points $x\in S$, but \textit{only those} $x \in S$ \textit{which fall outside the future light cone of} $y$. This modification will ultimately be responsible for the Lorentz covariant descriptions of quantum reality in Kent's relativistic framework, and will be covered in depth in section \ref{RelSolns}. 
 
%% ABL RULE
\newpage
\section{The Aharonov-Bergmann-Lebowitz Rule}
\label{ABL}
Aharonov, Bergmann, and Lebowitz (ABL) first derived in 1964 \cite{ABLpaper} what is now known as the ABL rule, and it plays a pivotal role in Kent's non-relativistic framework, with a very similar analog in the relativistic framework. To motivate the rule, consider a scenario in which an experimenter prepares a quantum system in a known initial state. Time passes, then a measurement is made of the quantum system, but the result of the measurement \textit{is not known to the experimenter}. More time passes, then the experimenter makes another, final measurement and finds the quantum system in some final state. Although the experimenter does not know the result of the intermediate measurement, but \textit{does} knows the initial and final states, can the experimenter at least determine the probabilities of possible results of the intermediate measurement?

Indeed the experimenter can. Using the ABL rule, knowledge of the final state provides further insight than is provided by the Born rule alone, and constrains the possible results of the intermediate measurement. Here the ABL rule is derived from basic principles in quantum theory and probability theory --- first in its simplest form, then generalized as necessary to suit Kent's framework. From here on, unless stated otherwise, natural units will be used so that $\hbar=c=1$. 

\subsection{The ABL Rule in Basic Form}
Consider a quantum state which is prepared in an eigenstate $\ket{a}$ of some observable $A$ at time $t=0$, and which is later measured and found to be in an eigenstate $\ket{c}$ of some observable $C$ at time $t=T$. What then is the probability of measuring some outcome $b_i$ of another observable $B$, measured in between $t=0$ and $t=T$, given that the state is initially in the state $\ket{a}$ and later measured and found to be in the state $\ket{c}$? 

In the most common and simplest case, the ABL rule is formulated for the case in which the eigenvalues of $B$ and $C$ are nondegenerate and discrete, and for which the Hamiltonian of the system is simply $H=0$, so that states persist unchanged between measurements. Beginning with these simplifications, the basic ABL rule will be derived then generalized as needed.

First, consider the probability of obtaining some result $b_i$ from the intermediate measurement followed by finding $c$ in the final measurement, given initial state $\ket{a}$. This is written as
\begin{equation}
\label{ABLFirstEqn}
\textrm{Pr}\left(\,b_i,\,c \bigm\vert a \right) = \textrm{Pr}\left(c\bigm\vert b_i , a\right) \cdot \textrm{Pr}\left(b_i\bigm\vert a\right)
\end{equation}
where terms like $\textrm{Pr}(x,\,y\bigm\vert u,\,v)$ denote the joint probability of $x$ and $y$ conditioned on $u$ and $v$. The chain rule for conditional probabilities was used to split the probability into two separate factors.

 With a Hamiltonian $H=0$, the system remains in the initial prepared state $\ket{a}$ up until the moment of measurement of $B$. The Born rule then prescribes the probability of obtaining outcome $b_i$ as $\textrm{Pr}\left(b_i\bigm\vert a\right)=\left|\bra{b_i}\ket{a}\right|^2$ with corresponding post-measurement state $\ket{b_i}$. Similarly, having measured $b_i$, the system remains in the state $\ket{b_i}$ up until the moment of final measurement of $C$ at time $T$. The probability of obtaining the result $c$, which depends only on the state $\ket{b_i}$ and not the initial state, is again prescribed by the Born rule as $\textrm{Pr}\left(c\bigm\vert b_i , a\right)=\textrm{Pr}\left(c\bigm\vert b_i \right)=\left|\bra{c}\ket{b_i}\right|^2$. Thus, equation (\ref{ABLFirstEqn}) can be evaluated as
\begin{equation}
\label{ABLSecondEqn}
\textrm{Pr}\left(\,b_i,\,c \bigm\vert a \right) = \left|\bra{c}\ket{b_i}\right|^2 \left|\bra{b_i}\ket{a}\right|^2.
\end{equation}
Now, the joint probability in (\ref{ABLFirstEqn}) may also be expanded in another way with the chain rule:
\begin{equation}
\label{ABLThirdEqn}
\textrm{Pr}\left(\,b_i,\,c \bigm\vert a \right) = \textrm{Pr}\left(\,b_i \bigm\vert c,a \right) \cdot \textrm{Pr}\left(\,c\bigm\vert a\right).
\end{equation}
Notice that the term $\textrm{Pr}\left(\,b_i \bigm\vert c,a \right)$, giving the probability of measuring $b_i$ given the initial and final states, is the one of interest for the ABL rule. Dividing both sides by $\textrm{Pr}\left(\,c\,\, |\, a\right)$ gives
\begin{equation}
\label{ABLUneval}
\textrm{Pr}\left(\,b_i \bigm\vert c,a \right) = \frac{\textrm{Pr}\left(\,b_i,\,c \bigm\vert a \right)}{\textrm{Pr}\left(\,c\bigm\vert a\right)}.
\end{equation}
The probability $\textrm{Pr}\left(\,c\,\, |\, a\right)$ can be interpreted as the total probability of ultimately measuring the eigenstate $\ket{c}$ given that the initial state was $\ket{a}$, and that observable B was measured in between. This may be evaluated using the law of total probability, accounting for every possible outcome of measuring B. Denoting by $\textbf{B}$ the set indexing all possible outcomes of the measurement of $B$, the total probability can be written as
\begin{equation}
\label{ABLFourthEqn}
\textrm{Pr}\left(\,c\bigm\vert a \right)=\sum_{j\in\textbf{B}}{\textrm{Pr}\left(\,c\bigm\vert b_j , a \right)\cdot \textrm{Pr}\left(\,b_j \bigm\vert a \right)}=\sum_{j\in\textbf{B}}{\textrm{Pr}\left(\,c \bigm\vert b_j \right)\cdot \textrm{Pr}\left(\,b_j \bigm\vert a \right)}.
\end{equation}
The conditional probability $\textrm{Pr}\left(\,b_i \bigm\vert c,a \right)$ in (\ref{ABLUneval}) can now be evaluated using equation (\ref{ABLSecondEqn}) and the total probability (\ref{ABLFourthEqn}) above. Thus, the ABL rule can be written in its simplest form as follows:
\begin{equation}
\textrm{Pr}\left(\,b_i\bigm\vert c , a \right) = \frac{\textrm{Pr}\left(\,c\bigm\vert b_i \right) \cdot \textrm{Pr}\left(\,b_i\bigm\vert a\right)}{\sum\limits_{j\in\textbf{B}}{\textrm{Pr}\left(\,c\bigm\vert b_j \right)\cdot \textrm{Pr}\left(\,b_j\bigm\vert a \right)}}=\frac{\left|\bra{c}\ket{b_i}\bra{b_i}\ket{a}\right|^2}{\sum\limits_{j\in\textbf{B}}{\left|\bra{c}\ket{b_j}\bra{b_j}\ket{a}\right|^2}}.
\end{equation}

\subsection{Degenerate Eigenvalues}
\label{DegenEigs}
The ABL rule can be generalized to accommodate eigenvalues of the measurements B and C which may be degenerate. It is no longer possible to uniquely label eigenstates like $\ket{b_i}$ or $\ket{c}$ with only their eigenvalue, as the eigenvalue $b_i$ may correspond to multiple eigenstates. Thus, another formalism for expressing probabilities like $\textrm{Pr}\left(\,b_i \bigm\vert a\right)$ is needed. In the more general formalism of projective measurements, the observable B is defined via a set of projection operators $\{P_{i}\}$ satisfying $P_{i}P_{j}=\delta_{i,j}P_{i}$ and the completeness relation $\sum_i{P_{i}}=I$, where the operator $P_i$ projects onto the eigenspace of the $i$'th outcome of the measurement. Given an arbitrary state $\ket{\phi}$ the probability of measuring the $i$'th outcome is given by $\textrm{Pr}\left(\,i \bigm\vert \ket{\phi}\right)=\bra{\phi}P_i\ket{\phi}$.

The post-measurement state after obtaining the $i$'th outcome of observable B must also be reconsidered, as it is no longer true that measuring outcome $i$ implies that the state collapses to a unique eigenstate. For this the L\"{u}ders rule is needed, which prescribes the post-measurement state of $\ket{\psi(t)}$ after obtaining the i'th outcome of observable B as $NP_i\ket{\psi(t)}$; where $N$ is a normalization constant to ensure that the state remains normalized. Thus, immediately after measurement of B, and still assuming the Hamiltonian is zero, the system is in the state $NP_i\ket{a}$.

The ABL rule can now be rewritten in the projective measurement formalism, which is valid for possibly degenerate eigenvalues of B and C. As above, $P_i$ is the projector onto the eigenspace corresponding to the i'th outcome of measurement B, and $P_c$ is the projector corresponding to the fixed outcome $c$ of measurement C.
\begin{equation}
\label{ABL_degen_1}
\textrm{Pr}\left(\,b_i\bigm\vert c , a \right) = \frac{\textrm{Pr}\left(\,c\bigm\vert b_i \right) \cdot \textrm{Pr}\left(\,b_i\bigm\vert a\right)}{\sum\limits_{j\in\textbf{B}}{\textrm{Pr}\left(\,c\bigm\vert b_j \right)\cdot \textrm{Pr}\left(\,b_j\bigm\vert a \right)}}=\frac{\big(\bra{a}NP_i\big)P_c\big(NP_i\ket{a}\big)\cdot \bra{a}P_i\ket{a}}{\sum\limits_{j\in\textbf{B}}{\big(\bra{a}NP_j\big)P_c\big(NP_j\ket{a}\big)\cdot \bra{a}P_j\ket{a}}}
\end{equation}

The expression may be put into a more compact form by rewriting the numerator above. Recall that $P_i$ is a projector, so that $P_i=P_iP_i$, and that $N$ is a scalar which we are free to commute about the expression. The numerator of (\ref{ABL_degen_1}) may then be written in the suggestive form
\begin{equation}
\Big(\bra{a}NP_i\Big)P_cP_i\ket{a}\bra{a}P_i\Big(NP_i\ket{a}\Big).
\end{equation}
Unit norm states $NP_i\ket{a}$ sandwich the expression. This allows one to write the above as a trace, using an orthonormal basis containing $NP_i\ket{a}$ and other orthogonal states, as any other basis state orthogonal to $NP_i\ket{a}$ will also be orthogonal to the $\bra{a}P_i$ term. Then
\begin{equation}
\textrm{Pr}\left(\,c\bigm\vert b_i \right) \cdot \textrm{Pr}\left(\,b_i\bigm\vert a\right)=\textrm{Tr}\big(P_cP_i\ket{a}\bra{a}P_i\big)=\textrm{Tr}\big(P_cP_iP_aP_i\big)
\end{equation}
where we denote by $P_a:=\dyad{a}$ the projector onto the initial state. Thus, the ABL rule can be rewritten in the compact form below, valid for degenerate eigenvalues of B and C:
\begin{equation}
\label{ABL_Degen}
\textrm{Pr}\left(\,b_i\bigm\vert c , a \right) = \frac{\textrm{Tr}(P_cP_iP_aP_i)}{\sum\limits_{j\in\textbf{B}}{\textrm{Tr}(P_cP_jP_aP_j)}}.
\end{equation}

\subsection{Time Evolution}
If the quantum state exhibits non-negligible time evolution between measurements, then there is a non-zero Hamiltonian $H$ which must be accounted for. Immediately before measurement of B at time $t$, the system will have unitarily evolved to the state $e^{-iHt}\ket{a}$. According to L\"{u}ders rule, immediately after measurement of B the system is then in the state $NP_ie^{-iHt}\ket{a}$. Following unitary evolution up until time $T$ immediately before measurement of C, the system is finally in the state $Ne^{-iH(T- t)}P_ie^{-iH t}\ket{a}$. The probabilities in ABL formula (\ref{ABL_Degen}) thus require a revision:
\begin{equation}
\begin{split}
\textrm{Pr}\left(\,b_i \bigm\vert a\right) &= \bra{a}e^{iH t}P_ie^{-iH t}\ket{a} \\
\textrm{Pr}\left(\,c \bigm\vert b_i\right) &= \big(\bra{a}Ne^{iH t}P_ie^{iH(T- t)}\big)P_c\big(Ne^{-iH(T- t)}P_ie^{-iH t}\ket{a}\big).
\end{split}
\end{equation}
From section \ref{DegenEigs}, the product of the probabilities above can be written compactly as a trace on the projectors.
\begin{equation}
\textrm{Pr}\left(\,c \bigm\vert b_i\right) \cdot \textrm{Pr}\left(\,b_i \bigm\vert a\right)=\textrm{Tr}(e^{iH(T- t)}P_ce^{-iH(T- t)}P_ie^{-iH t}P_ae^{iH t}P_i)
\end{equation}
The full ABL rule generalized for non-zero Hamiltonians thus takes the updated form
\begin{equation}
\label{ABL_sum}
\textrm{Pr}\left(\,b_i\bigm\vert c , a \right)= \frac{\textrm{Tr}(e^{iH(T- t)}P_ce^{-iH(T- t)}P_ie^{-iH t}P_ae^{iH t}P_i)}{\sum\limits_{j\in\textbf{B}}{\textrm{Tr}(e^{iH(T- t)}P_ce^{-iH(T- t)}P_je^{-iH t}P_ae^{iH t}P_j)}}\,.
\end{equation}
This is the form of the ABL rule used by Kent in his non-relativistic class of solutions, presented in section \ref{NonRel}. With this formula in hand, it is then straightforward to obtain expectation values for beables at intermediate times by averaging over all possible values weighted by their above ABL probability.

\subsection{In Relativistic Solutions}
\label{BackToBorn}
One may hope that the ABL rule extends to relativistic frameworks. As before, one could construct the probability of a given outcome $b_i$ at an intermediate spacetime point $y$ by pre-selecting on the initial condition on the initial hypersurface $S_0$ and post-selecting on the final boundary condition on the final hypersurface $S$. However, in general there are points on both $S_0$ and $S$ which are spacelike separated from $y$, and these points cannot be regarded as being in the past or future of $y$ in any meaningful sense. Thus, the ABL rule as originally envisaged, namely relating an initial state in the past of $y$ and a final state in the future of $y$, does not extend naturally.

Full specification of the conditional probability replacing the ABL rule in relativistic solutions for intermediate measurements of B requires further definitions which are postponed until the relativistic solutions are described in section \ref{RelSolns}. However, Kent will still make use of a conditional probability for an outcome $b_i$ given a relevant ``final" condition $c$ and an initial state $\ket{a}$. Formally, the conditional probability may still be expressed using equations (\ref{ABLThirdEqn}) and (\ref{ABLFourthEqn}) as
\begin{equation}
\textrm{Pr}\left(\,b_i\bigm\vert c , a \right) = \frac{\textrm{Pr}\left(\,c\bigm\vert b_i \right) \cdot \textrm{Pr}\left(\,b_i\bigm\vert a\right)}{\sum\limits_{j\in\textbf{B}}{\textrm{Pr}\left(\,c\bigm\vert b_j \right)\cdot \textrm{Pr}\left(\,b_j\bigm\vert a \right)}} .
\label{RelABL}
\end{equation}

Fortunately, the conditional probabilities are often highly simplified in the relativistic solutions. Upon postulating the existence of ``photon-like" particles in the system, which are treated as point-like particles moving along lightlike spacetime segments, there are only a finite and often small number of possible final boundary conditions and intermediate beable values to be considered. The conditional probabilities are reduced to something like a geometric game of ``photon billiards".

Recall from section \ref{KentStep4} that in the relativistic solutions, when considering the conditional probabilities at a spacetime point $y$, Kent post-selects not on the entire final boundary condition but only on that part outside the future light cone of $y$. This leads to an interesting case: what happens when none of the relevant final boundary condition is ``visible" outside the future light cone? 

More precisely, for a given intermediate spacetime point $y$, it is possible that every possible final boundary condition $\rho_i(x)$ is \textit{the same} for points $x$ which lie on the final hypersurface $S$ and outside the future light cone of $y$. That is, $\rho_i(x)=\rho_j(x)$ for all $x\in S$ outside the future light cone of $y$. This will often be the case, at least in our toy models to come, for spacetime points far in the past of the relevant interactions which take place in the system. At these points far in the past, none of the photons correlated with system are yet able to escape the future light cone, resulting in all of the final boundary conditions being indistinguishable outside the light cone. Now, only one of the final boundary conditions will correspond to the one actually realized in the quantum system. However, in this special case, every final boundary condition is equivalent for the set of points $x\in S$ outside the future light cone. Thus, no matter which final boundary condition is actually chosen, the result of fictitiously measuring the final quantum state outside the light cone on $S$ is certain to give just one possible outcome. Thus, in our notation using $c$ as the relevant final boundary condition, there is only one possible condition $c$, and it is certain. That is, in (\ref{RelABL}) above, $\textrm{Pr}\left(\,c\bigm\vert b_i \right)=1$ for all $b_i$. This leads to a drastic simplification of the conditional probability:
\begin{equation}
\textrm{Pr}\left(\,b_i\bigm\vert c , a \right) = \frac{ \textrm{Pr}\left(\,b_i\bigm\vert a\right)}{\sum\limits_{j\in\textbf{B}}{\textrm{Pr}\left(\,b_j\bigm\vert a \right)}}
\end{equation}

The denominator can be recognized as the total probability of measuring any one of the outcomes $b_i$ given the initial state $\ket{a}$. However, one is of course certain to obtain one of the outcomes after measurement of B, thus the denominator simply sums to one. 

In the numerator, all that remains is $\textrm{Pr}\left(\,b_i\bigm\vert a\right)$, the probability of measuring $b_i$ given the initial state $\ket{a}$. To evaluate this probability, first consider evolving the initial state forward in time to the point of measurement. For time evolution, Kent employs the Tomonaga-Schwinger formalism, which defines a unitary evolution operator between spacelike hypersurfaces. Considering a hypersurface in the future of $S_0$ which contains the spacetime point $y$ of interest, the Tomonaga-Schwinger formalism produces some state $\ket{a'}$ giving the unitarily evolved initial state on that hypersurface. 

Now, the probability of measuring a given outcome $b_i$ given the evolved state $\ket{a'}$ is prescribed by perhaps the most successful empirical law of quantum mechanics: the Born rule. Thus, there exists a complete reduction of the conditional probabilities used by Kent to the standard Born rule in the case when there is no relevant information outside the light cone. This observation and its implications are the foundation for a forthcoming analysis \cite{BRR} of Kent's interpretation.

%% NONRELATIVISTIC SOLUTIONS
\newpage
\section{Non-Relativistic Models}
\label{NonRel}

All of the tools needed to start applying Kent's interpretation to model systems are now available. One may imagine the non-relativistic framework would be simpler than the relativistic framework, with Minkowski spacetime structure and post-selection outside the light cone. However, it will become clear that in fact the opposite is closer to the truth. Nevertheless, let us first construct solutions to the quantum reality problem in non-relativistic three-dimensional Euclidean space. First, the most straightforward approach one might hope for will be taken to construct a solution to the quantum reality problem. It will ultimately fail, but motivate a successful second attempt and illuminate the main challenges.

\subsection{An $N$-Particle System}
\label{NParticles}
How far can Kent's interpretation be taken in describing the physical reality of a generic $N$-particle system? This idea was originally presented by Kent \cite{KentYoutube} in broad strokes but will be constructed here in detail. 

Consider an $N$-particle system governed by pure unitary evolution generated by a Hamiltonian $H$. Suppressing any intrinsic degrees of freedom, Kent considers mass as the preferred beable of the theory, allowing the $i$'th particle to have its own mass $m_i$. The quantum sample space for the system then consists of asymptotic final boundary conditions in the form of final mass distributions of the system. Kent's main postulate is then that upon random selection of a single final mass distribution from the quantum sample space, the expectation value of the mass density of the system may be defined for all points in spacetime.

As an initial condition, suppose the initial wave function of the system is provided as $\psi_0(\vec{x_1},\ldots,\vec{x_N})$ for all $\vec{x}_i\in \mathbb{R}^3$. The unitarily evolved wave function at any later time is then given by
\begin{equation}
\psi(\vec{x_1},\ldots,\vec{x_N};t) = e^{-iHt}\,\psi_0(\vec{x_1},\ldots,\vec{x_N})\,.
\end{equation}

At a late time $T$, which is taken in the limit at $T$ approaches infinity, it is postulated that there exists a final mass distribution $\rho(\vec{x};T)$. As explained in section \ref{Introduction}, this final boundary condition is mathematically described as the result of a fictional mass density measurement, here characterized by a mass weighted sum of position measurements. Spacelike separated position measurements will be assumed to commute, so that simultaneous position measurements at all points in space may be considered. Measuring the position of the $i$'th particle completely localizes the particle to some position $\vec{y_i}$, thus upon such a measurement the probability density of each particle's position collapses to the position eigenstate $\delta^3(\vec{x}-\vec{y_i})$. The form of the final boundary condition is then a mass weighted sum of probability densities
\begin{equation}
\label{NonRelLate}
 \rho(\vec{x};T) = \sum_i^N m_i \, \delta^3(\vec{x}-\vec{y_i})\,.
\end{equation}

Now to define expectation values for the mass density distribution at intermediate times $t$ and positions $\vec{x}$. To do so, the ABL rule will be applied given the initial wave function $\psi_0$ and final mass density distribution $\rho(\vec{x};T)$. Following section \ref{ABL}, the ABL rule probability of measuring some mass $M_i$ in a small volume $\delta V$ around point $\vec{x}$ at time $t$ may be computed. Taking the limit as $\delta V$ goes to zero then produces the mass density at point $\vec{x}$. The masses $M_i$ are labelled with subscript $i$ as there are in fact only a finite number of possible masses which may be measured. Recall mass density measurements are defined via a simultaneous measurement of particle position at all points in space, localizing all particles to positions $\{y_i\}$, then mass weighting the delta functions as in (\ref{NonRelLate}) above. Thus, upon measurement, the entire mass of the $i$'th particle either will or will not be contained within the small volume $\delta V$. The only possible outcomes of a mass measurement are then any combination of the $m_i$'s summed together, which, for finite $N$, is a finite set of masses. 

Precise language will be needed in order to properly set up the ABL probability. For clarity below, all boldfaced letters correspond to sets. Let $\textbf{N}=\{1,\ldots,N\}$ and define the set $\textbf{K}=\{ \textbf{k}\subseteq\textbf{N}  :   \textbf{k}\neq\emptyset \}$ as the set of all non-empty subsets of $\textbf{N}$. Each set $\textbf{k}\in \textbf{K}$ corresponds to a unique subset of the $N$ particles. The combined mass of this collection of particles is then $M_{\textbf{k}} = \sum_{i\in \textbf{k}} m_i $. Finally let the set $\textit{\textbf{M}}= \{M_{\textbf{k}}  : \textbf{k} \in \textbf{K}  \}$ be the finite set of all such possible combined masses.

The various projection operators appearing in the ABL probability must now be formulated. Let $P_0=\dyad{\psi_0}$ be the projector onto the initial state, and $P_F=\dyad{\vec{y_1},\ldots,\vec{y_N}}$ be the projector onto the final state with mass distribution (\ref{NonRelLate}). The projector at intermediate time $t$ requires a bit more work; the projector $P_{M_i}^{\vec{x}}$ is required to project onto the eigenspace of states with mass $M_i \in \textbf{M}$ in the volume $\delta V$ around a point $\vec{x}$, where $\delta V$ will tend to zero. This requires that summing over all collections of particles $\textbf{k}$ which have combined mass $M_{\textbf{k}}=M_i$. For any such collection $\textbf{k}$, all particles $i\in\textbf{k}$ must be projected onto the space inside $\delta V$ but allow the other particles to assume any position in space by integrating over all such possibilities. For notational convenience, define the set $\textbf{k}^C$ as the complement of $\textbf{k}$ with respect to $\textbf{N}$, containing all particles not in $\textbf{k}$. 

Define the projectors $P_j^{\vec{y}}$ to project the system onto the space of states with the $j$'th particle localized at position $\vec{y}$. As the $P_j^{\vec{y}}$ project the $j$'th particle into a position eigenstate, it necessarily follows that $P_j^{\vec{y}}P_j^{\vec{x}}=\delta^3(\vec{x}-\vec{y})P_j^{\vec{y}}$. One may then construct a projector $P_{M_i}^{\delta V}$ onto the space of states with mass $M_i$ in the volume $\delta V$ as follows. Below, $\textrm{d}^3\vec{x_{\textbf{k}}}=\prod_{j\in\textbf{k}}\textrm{d}^3\vec{x_j}$ and $\textrm{d}^3\vec{x}_{\textbf{k}^C}=\prod_{l\in\textbf{k}^C}\textrm{d}^3\vec{x_l}$ are written as a shorthand for the measures over positions.
\begin{equation}
\label{bigprojector}
 P_{M_i}^{\delta V} \coloneqq \sum_{\textbf{k} \, | \, M_{\textbf{k}}=M_i} \int_{\delta V} \textrm{d}^3\vec{x_{\textbf{k}}} \prod_{j \in \textbf{k}} P_j^{\vec{x_j}} \int_{\mathbb{R}^3\setminus \delta V} \textrm{d}^3\vec{x}_{\textbf{k}^C} \,\prod_{l \in \textbf{k}^C} P_l^{\vec{x_l}}
 \end{equation}
However, in the limit as the volume of $\delta V$ tends to zero around $\vec{x}$, when computing the probability density, the only point which remains inside $\delta V$ is $\vec{x}$. Thus define the projector $P_{M_i}^{\vec{x}}$, projecting onto the space of states with mass $M_i$ in the differential volume d$V$ around the position $\vec{x}$, as
\begin{equation}
P_{M_i}^{\vec{x}}\,\textrm{d}V \coloneqq \sum_{\textbf{k} \, | \, M_{\textbf{k}}=M_i}  \int_{\mathbb{R}^3} \textrm{d}^3\vec{x}_{\textbf{k}^C} \, \prod_{j \in \textbf{k}} P_j^{\vec{x}}\prod_{l \in \textbf{k}^C} P_l^{\vec{x_l}} \,.
\end{equation}

In the limit where the volume $\delta V$ becomes infinitesimal around $\vec{x}$, presumably only one particle at most will ever sit inside $\delta V$, though such details would be coded into the given Hamiltonian $H$ defining particle interactions. In the case that it is indeed true that two particles cannot both be perfectly localized to the same point in space after a position measurement, the projector $P_{M_i}^{\vec{x}}$ above can be written as
\begin{equation}
\label{NoOverlapProjector}
P_{M_i}^{\vec{x}} \, \textrm{d}V = \sum_{j \, | \, m_j=M_i} \int \textrm{d}^3\vec{x_1} \ldots \textrm{d}^3\vec{x}_{j-1}\textrm{d}^3\vec{x}_{j+1}\ldots\textrm{d}^3\vec{x_N} \, P_j^{\vec{x}} \prod_{k\neq j}P_k^{\vec{x_k}} \,.
\end{equation}

As the next step in writing down the ABL probability, one must construct the total probability term $\textrm{Pr}(\,\rho(\vec{x};T) \, |\, \psi_0 )$ expressing the probability of measuring the late-time mass distribution $\rho(\vec{x};T)$ given the initial wave function $\psi_0$. This probability is expressed by considering all possible intermediate measurements of mass $M_i$ at any position and summing over all possible mass measurements $M_i$. To this end, define the projector $P_{M_i}$ which projects onto the space of states with mass $M_i$ at any point.
\begin{equation}
P_{M_i} \coloneqq \int_{\mathbb{R}^3} \textrm{d}^3\vec{x}\,P_{M_i}^{\vec{x}}
\end{equation}
With a slew of projectors defined, the ABL rule (\ref{ABL_sum}) can be evaluated for the probability density of measuring mass $M_i$ at the point $\vec{x}$ and time $t$, given initial state $\ket{\psi_0}$ and final mass distribution $\rho(\vec{x};T)$. 
\begin{equation}
\label{NonRelABL}
\textrm{Pr}\left(\,M_i,\vec{x},t\bigm\vert \psi_0 \,,\, \rho(\vec{x};T)\right) = \frac{\textrm{Tr}(e^{iH(T-t)}P_Fe^{-iH(T-t)}P_{M_i}^{\vec{x}}e^{-iHt}P_0e^{iHt}P_{M_i}^{\vec{x}})}{\sum\limits_{M_j\in \textbf{M}}{\textrm{Tr}(e^{iH(T-t)}P_Fe^{-iH(T-t)}P_{M_j}e^{-iHt}P_0e^{iHt}P_{M_j})}}
\end{equation}
where all of the projectors $P_0$, $P_F$, $P_{M_i}^{\vec{x}}$, and $P_{M_j}$ are defined above but left unexpanded here for readability. Finally, the expectation value of the mass density is obtained by averaging over all $M_i$ weighted by the corresponding probability of $M_i$ directly above.
\begin{equation}
\label{ExpectationValue}
\langle\, \rho(\vec{x};t) \, \rangle = \sum_{M_i \in \textbf{M}} M_i \cdot  \textrm{Pr}\left(\,M_i,\vec{x},t\bigm\vert \ket{\psi_0} \,,\, \rho(\vec{x};T)\right)
\end{equation}
Viol\`a, it would seem that Kent's framework has succeeded in at least formally defining expectation values of mass density at all points in spacetime, even if the solution is far from explicit. Unfortunately, although this formula gives an answer, the answer is not a good one. In Kent's own words, but which was also known quite well by Heisenberg, the problem arises because localizing a particle to a single point ``is a very violent thing to do to it" \cite{KentYoutube}. 

Let's understand how the problem arises. Maintaining a non-relativistic point of view, the Heisenberg uncertainty principle says that complete localization of a particle in position space results in complete delocalization in momentum space. When considering a mass-weighted position measurement in the ABL rule at an intermediate time, the particles collapse into position eigenstates, which are of course delta functions around the measured particle positions $\{z_i\}$ in position space.
\begin{equation}
\psi(\vec{x_1},\ldots,\vec{x_N};t) \to \prod_i^N \delta^3(\vec{x_i}-\vec{z_i}) 
\end{equation}
In Fourier space, the wave function is thus completely delocalized, taking the form
\begin{equation}
\mathcal{F}[ \psi(\vec{x_1},\ldots,\vec{x_N};t) ] \to \int \textrm{d}^3\vec{x}_{\textbf{N}}\,\, \prod_i^N  \delta^3(\vec{x_i}-\vec{z_i})\,e^{-2\pi i \, \vec{x_i} \cdot \vec{\omega_i}}= \prod_i^N e^{-2\pi i \, \vec{z_i} \cdot \vec{\omega_i}}
\end{equation}
with non-normalizable momentum space probability density $|\psi(\vec{\omega_1},\ldots,\vec{\omega_N};t)|^2 = 1$. Thus, after the intermediate position measurement, the particles may obtain arbitrarily high momentum. The effect is that \textit{any} set of particle positions at the late time $T$ \textit{are equiprobable} after the intermediate position measurement, and the final boundary condition is completely uncorrelated with the intermediate time positions.

Thus, the formally correct expression (\ref{NonRelABL}) for the ABL probability $\textrm{Pr}(\,M_i,\vec{x},t\,\, |\, \psi_0 \,,\, \rho(\vec{x};T))$ of measuring mass $M_i$ at the point $\vec{x}$ and time $t$ in fact assumes a much simpler form. For any given mass $M_i$, the corresponding probability of measuring $M_i$ is simply the number of particles with mass $M_i$ over the total number of particles.
\begin{equation}
\label{catastrophe}
\textrm{Pr}(\,M_i,\vec{x},t\,\, |\, \psi_0 \,,\, \rho(\vec{x};T)) = \sum_{j\,|\,m_j=M_i} \frac{1}{N}
\end{equation}

The ABL probabilities are thus constant over all spacetime as given above, leading to an expectation value (\ref{ExpectationValue}) which is also constant over all spacetime. Thus, our beables, the expectation values of mass density in spacetime, are constant throughout our universe. An odd result which does not intuitively describe observed quasiclassical reality.

This catastrophe indicates that a particle's position alone at late times is not enough to come up with any non-trivial or intuitively appealing statement regarding expectation values of its mass density at earlier positions. Before making another attempt, one might first argue that this attempt may not have been for nought if a relativistic attitude had been taken; since after the intermediate position measurement, one can be sure that our wild particles at least stay within their future light cone. One would be correct, and it is for this reason and others that Kent's framework becomes more natural in relativistic settings---as will be seen soon enough.

\subsection{Interacting Classes of Particles}
Suppose you decide to set two billiard balls on a crash course for each other on a frictionless infinite plane, with two initially known positions and velocities. Measuring the position of one billiard ball well after the collision tells you quite a lot about the other - in fact, from conservation of momentum one could ascertain exactly the position of the other. Generalizing to a larger set of billiard balls, the picture gets more complicated but the general idea remains intact - knowledge of the position of some balls yields much information about the others. Kent builds from this sort of intuition, in a much more sophisticated way, to improve on his first prototype model described in section \ref{NParticles}.

Consider a further postulate that there exists a natural classification of particles into interacting classes. As Kent introduces in \cite{MainKent}, it is natural to assume two interacting classes composed of indistinguishable fermions and bosons, and also allow for other distinguishable particles in either class. By allowing the two classes of particles to interact for a finite time and assuming that the particles asymptotically reach well separated non-interacting final states, the final mass distribution of the class 1 particles can be used to post-select on expectation values for the class 2 particles' mass densities at intermediate times, and vice versa.

The previous model will require some modification but will remain largely intact. Again consider an $N$-particle system, with $b\geq2$ indistinguishable bosons, $f\geq2$ indistinguishable fermions, and $d\geq0$ distinguishable particles such that $b+f+d=N$. When labelling particles, let the first $\{1,\ldots,b\}$ label the indistinguishable bosons, the next $\{b+1,\ldots,b+f\}$ label the indistinguishable fermions, and the final $\{b+f+1,\ldots,N\}$ label the distinguishable particles.  Now designate two classes of particles: class B composed of all the bosons and class F composed of all the fermions, with each distinguishable particle allocated to either class B or F. Again letting $\textbf{N}=\{1,\ldots,N\}$, define the sets $\textbf{B}=\{1,\ldots,b\}\cup\{i>b+f:\textrm{particle } i \in \textrm{class B}\}$ and $\textbf{F}=\{b+1,\ldots,b+f\}\cup\{i>b+f:\textrm{particle } i \in \textrm{class F}\}$ containing the labels of the particles in each class, so that $\textbf{B}\cup\textbf{F}=\textbf{N}$. 

The mass of the $i$'th particle is again labelled as $m_i$, though it is required that all bosons have mass $m_B$ and all fermions have mass $m_F$. In this model it is assumed from the start that particles may not perfectly overlap in space after a position measurement, an assumption that, in the previous model, was only made later in equation (\ref{NoOverlapProjector}). This assumption entails that the result of a mass measurement in space may only yield one of the individual $m_i$'s, as opposed to any sum of them. Then define the sets of possible mass measurements for each class as $\textbf{M}_B=\{m_i \,|\, i \in \textbf{B}\}$ and $\textbf{M}_F=\{m_i \,|\, | i \in \textbf{F}\}$, noting that $\textbf{M}_B$ contains the possibly different masses of the distinguishable particles in class $\textbf{B}$ as well as $m_B$, and similarly for $\textbf{M}_F$.

Again suppose the initial quantum state $\ket{\psi_0}$ is given. By the new postulate, it is supposed that the composite state of the system $\ket{\psi(t)}$ decomposes into a tensor product of the individual states $\ket{\psi_1(t)}\otimes\cdots\otimes\ket{\psi_N(t)}$. The wave function of the system is then $\psi(\vec{x_1},\ldots,\vec{x_N};t) = \bra{\vec{x_1},\ldots,\vec{x_N}}\ket{\psi(t)}$ and note that the wave function must obey the proper spin statistics, so that $\psi(\vec{x_1},\ldots,\vec{x_N};t)$ is symmetric under exchange of any two bosonic labels and antisymmetric under exchange of any two fermionic labels. Time evolution of states is again provided by a Hamiltonian $H$ which defines particle interactions between and within classes, so that the state evolves as $\ket{\psi(t)}=e^{-iHt}\ket{\psi_0}$. 

An essential feature of this interacting class model is that in the ABL probability of measuring a given mass for the class B particles, only the final mass distribution of the class F particles are post-selected on (and vice versa). Thus, late-time mass measurements not of the entire system, but of just one class, must be considered. As before, late time $T$ mass distributions are constructed through a mass-weighted sum of position measurements. Position measurements are performed as before, with the operators $P_i^{\vec{x}}$ projecting onto the space of states with particle $i$ localized at position $\vec{x}$. Given the tensor product structure of the state, one may now write these projectors as
\begin{equation}
\label{PositionProjector}
 P^{\vec{x}}_i = I_1 \otimes \cdots \otimes I_{i-1} \otimes \ket{\vec{x}}_i \bra{\vec{x}}_i \otimes I_{i+1} \otimes \cdots \otimes I_N \,.
 \end{equation}
These projectors may then be used to construct mass density functions for a given particle. Denote $\rho_i(\vec{x};t)$ as the mass density function for the $i$'th particle at position $\vec{x}$, which, in words, gives the total probability of measuring this $i$'th particle at position $\vec{x}$ weighted by its mass $m_i$:
\begin{equation}
\begin{split}
 \rho_i(\vec{x};t) & = m_i \bra{\psi(t)}P^{\vec{x}}_i\ket{\psi(t)} \\
 & = m_i \int \textrm{d}^3\vec{x_1}\cdots\textrm{d}^3\vec{x_{i-1}}\textrm{d}^3\vec{x_{i+1}}\cdots\textrm{d}^3\vec{x_N}\,\,|\psi(\vec{x_1},\ldots,\vec{x_{i-1}},\vec{x},\vec{x_{i+1}},\ldots\vec{x_{N}};t)|^2 \,. \\
 \end{split}
 \end{equation}
It is worth noting that $P^{\vec{x}}_i$, and thus $\rho_i(\vec{x};t)$, is not well-defined on its own if $i$ is an indistinguishable bosonic or fermionic label in the set $\{1,\ldots,b+f\} \subseteq \textbf{N}$. As the bosons and fermions are indistinguishable, how could one be sure that $P^{\vec{x}}_i$ projects the $i$'th boson and not the $j$'th boson? All that one can be sure of is that $P^{\vec{x}}_i$ projects \textit{one} of the bosons onto position $\vec{x}$. However, sense can be made of operations including all of the bosons, such as $\sum_{i=1}^{b}P^{\vec{x}}_i$, which projects all of the bosons onto position $\vec{x}$. Similarly, this lets one place the mass density function of all the bosons $\sum_{i=1}^{b}m_i\bra{\psi(t)}P^{\vec{x}}_i\ket{\psi(t)}$ on solid footing. The same story holds for the fermions, summing over $i=b+1,\ldots,b+f$ instead. 
 
Importantly, the above lets one define the mass density functions for the two classes. The collective mass density functions above for all of the bosons or fermions is well defined, and the addition of any number of distinguishable particles presents no further problem. Thus, the mass density functions $\rho_B(\vec{x};t)$ and $\rho_F(\vec{x};t)$ may be defined, giving the probability density of finding a class B or class F particle at position $\vec{x}$ at time $t$, weighted by the mass of the particle:
\begin{align}
\rho_B(\vec{x};t) = \sum_{i\in \textbf{B}} m_i\bra{\psi(t)}P^{\vec{x}}_i\ket{\psi(t)} && \rho_F(\vec{x};t) = \sum_{i\in \textbf{F}} m_i\bra{\psi(t)}P^{\vec{x}}_i\ket{\psi(t)} \,.
\end{align}

From the above the total mass density of the system can constructed as in the first model in section \ref{NParticles} by simply adding the mass densities of the two classes: $\rho(\vec{x};t)=\rho_B(\vec{x};t)+\rho_F(\vec{x};t)$. 

The final boundary must now be considered, again characterized as a fictitious late time $T$ measurement of the mass density at all points in space. This yields a final mass distribution for post-selection in the ABL rule probability. However, to consider a simultaneous measurement at all points in space one need ensure that the operators $P^{\vec{x}}_i$ commute for all $\vec{x}$ and particle labels $i$, something taken for granted in section \ref{NParticles}. Without loss of generality, consider two such projectors $P^{\vec{x}}_i$ and $P^{\vec{y}}_j$ with $i\leq j$, both considered at an arbitrary fixed time $t$. In the case that $i\neq j$, using equation (\ref{PositionProjector}), the operators leave each other well alone.
\begin{equation}
P^{\vec{x}}_i P^{\vec{y}}_j = I_1 \otimes \cdots \otimes I_{i-1} \otimes \ket{\vec{x}}_i \bra{\vec{x}}_i \otimes I_{i+1} \otimes \cdots \otimes I_{j-1} \otimes \ket{\vec{y}}_j \bra{\vec{y}}_j \otimes I_{j+1} \otimes \cdots \otimes I_N = P^{\vec{y}}_j P^{\vec{x}}_i
\end{equation}
And for the case $i=j$, again referring to (\ref{PositionProjector}), one has
\begin{equation}
\setlength{\jot}{7pt}
\begin{split}
 P^{\vec{x}}_i P^{\vec{y}}_i & = I_1 \otimes \cdots \otimes I_{i-1} \otimes \ket{\vec{x}}_i \braket{\vec{x}}{\vec{y}} \bra{\vec{y}}_i \otimes I_{i+1} \otimes \cdots \otimes I_N \\
 & =I_1 \otimes \cdots \otimes I_{i-1} \otimes \ket{\vec{x}}_i \delta^3({\vec{x}}-{\vec{y}}) \bra{\vec{y}}_i \otimes I_{i+1} \otimes \cdots \otimes I_N \\
  & = I_1 \otimes \cdots \otimes I_{i-1} \otimes \ket{\vec{y}}_i \braket{\vec{y}}{\vec{x}} \bra{\vec{x}}_i \otimes I_{i+1} \otimes \cdots \otimes I_N \\
  & = P^{\vec{y}}_i P^{\vec{x}}_i\,.  \\
 \end{split}
 \end{equation}
Thus the commutator $[P^{\vec{x}}_i,P^{\vec{y}}_j]$ vanishes in all cases. Thus, simultaneous measurements of $P^{\vec{x}}_i$ for all points ${\vec{x}}$ in space are valid. In particular, simultaneous mass density measurements of the entire system can be defined via 
\begin{equation}
\sum_{i\in\textbf{N}} m_i\bra{\psi(t)}P^{\vec{x}}_i\ket{\psi(t)}
\end{equation}
As before, a fictitious measurement of the position of all particles localizes each particle $i$ to a delta function around some point $\vec{y_i}$, producing a late-time mass density distribution of the form
\begin{equation}
\begin{split}
\rho(\vec{x};T) & = \rho_B(\vec{x};T)+\rho_F(\vec{x};T) \\
& =  \sum_{i\in \textbf{B}} m_i\bra{\psi(T)}P^{\vec{x}}_i\ket{\psi(T)}+\sum_{i\in \textbf{F}} m_i\bra{\psi(T)}P^{\vec{x}}_i\ket{\psi(T)}\\
& = \sum_{i\in \textbf{N}} m_i\int \textrm{d}^3\vec{x_1}\cdots\textrm{d}^3\vec{x_{i-1}}\textrm{d}^3\vec{x_{i+1}}\cdots\textrm{d}^3\vec{x_N}\,\,|\psi(\vec{x_1},\ldots,\vec{x_{i-1}},\vec{x},\vec{x_{i+1}},\ldots\vec{x_{N}};T)|^2 \\
& = \sum_{i\in \textbf{N}} m_i\int \textrm{d}^3\vec{x_1}\cdots\textrm{d}^3\vec{x_{i-1}}\textrm{d}^3\vec{x_{i+1}}\cdots\textrm{d}^3\vec{x_N}\,\,\delta^3(\vec{x}-\vec{y_i})\prod_{j\neq i}\delta^3(\vec{x_j}-\vec{y_j})\\
& = \sum_{i\in \textbf{N}} m_i\,\delta^3(\vec{x}-\vec{y_i})\\
\end{split}
\end{equation}
And similarly, the class B and class F late-time mass distributions take the same form but summed only over their respective particles.
\begin{align}
\rho_B(\vec{x};T) = \sum_{i\in \textbf{B}}m_i \,\delta^3(\vec{x}-\vec{y_i}) && \rho_F(\vec{x};T) = \sum_{i\in \textbf{F}} m_i\,\delta^3(\vec{x}-\vec{y_i})
\end{align}

Now the ABL probabilities are constructed. Doing so is mostly a matter of carefully specifying the correct set of projection operators as before. For notational convenience and to avoid double quoting every equation henceforth, let $J$ be a placeholder for either class B or class F, and let $\bar J$ denote the other class, as the classes play symmetric roles. Similarly, define the sets $\textbf{J}$ and $\overline{\textbf{J}}$ containing the labels for particles in set $J$ and $\bar J$, respectively. Finally, the sets $\textbf{M}_J$ and $\textbf{M}_{\bar{J}}$ contain the possible outcomes of mass measurements for each class as defined earlier.

Consider the ABL probability for measuring a class $J$ mass of $M_i\in\textbf{M}_J$ at a position $\vec{x}$ and intermediate time $t$. The crux of the matter is that to evaluate the ABL probability for the class $J$ mass density, \textit{we post-select only on the late-time mass distribution of class }$\bar J$\textit{, the other class}. Thus, the projector $P_F$ appearing in the ABL probability only projects onto the space of states with the class $\bar J$ particles localized at their respective final positions:
\begin{equation}
P^{\bar J}_F \coloneqq \prod_{i\in\overline{\textbf{J}}}P_i^{\vec{y_i}}
\end{equation}

The projector $P_0$ onto the initial state remains untouched as $\dyad{\psi_0}$. A projector $P_{J,M_i}^{\vec{x}}$ is needed at an intermediate time $t$ to project onto the space of states which have a class $J$ particle of mass $M_i$ in the differential volume d$V$ at position $\vec{x}$. The positions of all other particles, in either class, are free to vary and are integrated over. As shorthand for the measures, we write $\textrm{d}^3\vec{x_{\textbf{N}}}=\prod_{j\in\textbf{N}}\textrm{d}^3\vec{x_j}$ and $\textrm{d}^3\vec{x_{\textbf{N}}}_{\setminus \{j\}}=\prod_{k\in\textbf{N} \,| k\neq j}\textrm{d}^3\vec{x_k}$, then define $P_{J,M_i}^{\vec{x}}$ via
\begin{equation}
P_{J,M_i}^{\vec{x}}\textrm{d}V \coloneqq \sum_{j\in\textbf{J} \, | \, m_j=M_i } \int \textrm{d}^3\vec{x_{\textbf{N}}}_{\setminus \{j\}} \, P_j^{\vec{x}} \prod_{k\in\textbf{N} \,:\, k\neq j}P_k^{\vec{x_k}}\end{equation}

The total probability term in the denominator of the ABL rule (\ref{ABL_sum}) requires a projector $P_{J,M_i}$, which, like the above, projects onto the space of states with a class $J$ particle of mass $M_i$, but allows the particle to be found anywhere by integrating over all positions of the particle. One then sums over all possible masses $M_i$ in the ABL probability.
\begin{equation}
P_{J,M_i} \coloneqq \int_{\mathbb{R}^3} \textrm{d}^3\vec{x}\,\,P_{J,M_i}^{\vec{x}}
\end{equation}

All of the pieces are in place. The ABL probability density for measuring a class $J$ mass of $M_i\in\textbf{M}_J$ at a position $\vec{x}$ and intermediate time $t$ is given below.
\begin{equation}
\textrm{Pr}(\,M_i,J,\vec{x},t\,\, |\, \psi_0 \,,\, \rho_{\bar J}(\vec{x};T)) = \frac{\textrm{Tr}(e^{iH(T-t)}P^{\bar J}_Fe^{-iH(T-t)}P_{J,M_i}^{\vec{x}}e^{-iHt}P_0e^{iHt}P_{J,M_i}^{\vec{x}})}{\sum_{M_j\in \textbf{M}}{\textrm{Tr}(e^{iH(T-t)}P^{\bar J}_Fe^{-iH(T-t)}P_{J,M_j}e^{-iHt}P_0e^{iHt}P_{J,M_j})}}
\end{equation}

Finally the expectation value of the mass density for class J at position $\vec{x}$ and time $t$ is obtained by averaging over all $M_i\in\textbf{M}_J$ weighted by the corresponding probability above.
\begin{equation}
\langle\, \rho_J(\vec{x};t) \, \rangle = \sum_{M_i \in \textbf{M}_J} M_i \cdot  \textrm{Pr}(\,M_i,J,\vec{x},t\,\, |\, \psi_0 \,,\, \rho_{\bar J}(\vec{x};T))
\end{equation}

Letting $J$ run over the bosonic and fermionic classes, the above defines the expectation values for the beables of our theory, the mass density of the fermionic and bosonic particles. In the quantum system in which all physics take place between times $t=0$ to $T$, with initial quantum state $\ket{\psi_0}$, quasiclassical physical reality is described by the following two generalized mass density fields on spacetime.
\begin{equation}
	\{\,\rho_T^B(\vec{x};t) \,,\, \rho_T^F(\vec{x};t) : 0<t<T \,,\, \vec{x}\in\mathbb{R}^3 \}
\end{equation}

A solution to the non-relativistic quantum reality problem may now be claimed. Having formally solved the quantum reality problem, some natural questions may arise, addressed now. 

\vspace{2mm}
\begin{center}
\textit{How does this solution escape the trivial description encountered in the previous model?}
\end{center}
\vspace{2mm}

Recall how the problem arose in the first model. The question being asked in the ABL rule probability was: given the position of all particles at an intermediate time, what is the probability of finding them in the correct final configuration at late time? The answer was trivial, as completely localizing the particles at the intermediate time destroys any ability to predict their future positions, as seen in equation (\ref{catastrophe}).

The key difference here is that, to derive an ABL probability of finding the fermions in some position at intermediate time, one considers an intermediate time measurement of the fermions positions, but then does not post-select on these fermion positions, only the bosons' positions. Thus, while any power to predict future fermion positions is ruined by measuring their positions at an intermediate time, \textit{ones does not lose power to predict boson positions after measuring the fermions}, as the bosons themselves \textit{were not} measured at an intermediate time. The same story of course holds, replacing bosons with fermions and vice-versa. Thus, assuming there are non-trivial interactions between the bosons and fermions, the ABL probabilities are now non-trivial as well.

\vspace{2mm}
\begin{center}
\textit{Is this solution unique?}
\end{center}
\vspace{2mm}

No. Kent's framework can accommodate different beables for the theory, and Kent himself considers alternatives such as the electromagnetic field in later work \cite{PhysRevA.96.062121}.

\vspace{2mm}
\begin{center}
\textit{Which is the correct solution then?}
\end{center}
\vspace{2mm}

One might suggest that, since there appear to be numerous ways of defining physical reality via different beables, only one such description should be correct. Or, perhaps, a single beable may not be enough to describe a ``full" physical reality. Does there then exist a fundamental set of beables which collectively describe quantum reality through their expectation values in spacetime? These are interesting questions to be explored!

%% RELATIVISTIC SOLUTIONS
\newpage
\section{Semi-Relativistic Models}
\label{RelSolns}
The principles of relativity suggest that the spacetime manifold of the universe, at least locally, looks like Minkowski space. Any interpretation of quantum theory would then ideally extend naturally to Minkowski space and respect the symmetries of special relativity. This is not always the case --- it has proven challenging to extend other interpretations addressing the quantum reality problem, such as de Broglie-Bohm pilot wave theory \cite{PilotWaveLorentz}, to Lorentz-covariant interpretations. However, Kent has proposed a relativistic interpretation, modifying his interpretation considered in section \ref{NonRel}, which describes physical reality using fully Lorentz-covariant rules, appropriate for Minkowski space and other background spacetimes. 

Intriguingly, Schr\"odinger's cat might just escape quantum purgatory in Kent's relativistic interpretation. As will be seen, the relativistic form of Kent's interpretation paints an interesting picture of physical reality for quantum states in superposition. While all kinds of particles or field perturbations are treated on equal footing, photon-like particles moving at the speed of light play an important preferred role, acting much like an ``environment" on which the quantum state leaves a physical mark. Through interaction between the quantum state and these photons, the beables of systems in superpositions may be knocked into definite configurations localized in space, producing a Lorentz-covariant ``collapse" of the beable values. That is, while the quantum state itself persists in a superposition, the physical beables describing reality may not reflect that superposition, instead assuming a configuration corresponding to a measurement eigenstate. 

First, the basic structures involved in Kent's nonrelativistic framework, presented in section \ref{NonRel}, will be extended to their Lorentz-covariant counterparts in subsection A. In the following two subsections, B and C, illuminating semi-relativistic toy models will be considered, with interactions between photon-like particles and larger quantum systems which are initially in superpositions of spatially well-separated states.

%% GENERAL FORM OF RELATIVISTIC SOLUTIONS 
\subsection{The General Form}
The general structure of Kent's framework remains mostly intact. The spirit of the ABL rule retains a prominent role, though a key modification is made which is responsible for the Lorentz covariant description of physical reality in the relativistic framework. Some new constructions are needed to properly express the new form of the conditional probabilities for intermediate measurements.

An initial quantum state $\ket{\psi_0}$ is now more generally provided on some spacelike hypersurface $S_0$. In the non-relativistic setting, the unitary evolution of quantum states was generated by a Hamiltonian operator, from an initial time $t=0$ to a later time $T$. In the relativistic setting, the generalization of the evolution of quantum states between arbitrary spacelike surfaces in Minkowski space is provided by the Tomonaga-Schwinger formalism \cite{TomonagaSchwinger}. Given the initial state $\ket{\psi_0}$ on $S_0$, the Tomonaga-Schwinger formalism provides a unitary operator $U_{SS_0}$ which gives the evolved state of the system $U_{SS_0}\ket{\psi_0}$ on any hypersurface $S$ in the future of $S_0$.

Now it is postulated that there exists a single final boundary condition, a single final beable configuration on the final hypersurface $S$, chosen by Nature from the sample space. Recall that a convergent sequence of ever later hypersurfaces is used to define the final boundary conditions, each with beable probability density prescribed by the quantum state as evolved up to that hypersurface. As a shorthand for the limiting hypersurface of this sequence, let us simply consider a single hypersurface $S$ with evolved state $U_{SS_0}\ket{\psi_0}$. 

To be explicit, let us consider mass-energy density as the beable, where it is assumed that spacelike measurements of mass-energy commute.  First, define the mass-energy density distribution on $S$ as $T_S(x)=T_{\mu\nu}(x)\hat n^\mu(x)\hat n^\nu(x)$, where $T_{\mu\nu}(x)$ is the stress-energy tensor at the point $x\in S$ and $\hat n(x)$ is the forward pointing timelike unit 4-vector normal to the tangent plane of $S$ at $x$. Then a single such distribution $t_S(x)$ is chosen by Nature which corresponds to physical reality. Keep in mind that no physical measurement of $T_S(x)$ is actually performed by any kind of outside observer, though we mathematically characterize $t_S(x)$ as the result of such.

\begin{figure}[h!]
\centering
\includegraphics[]{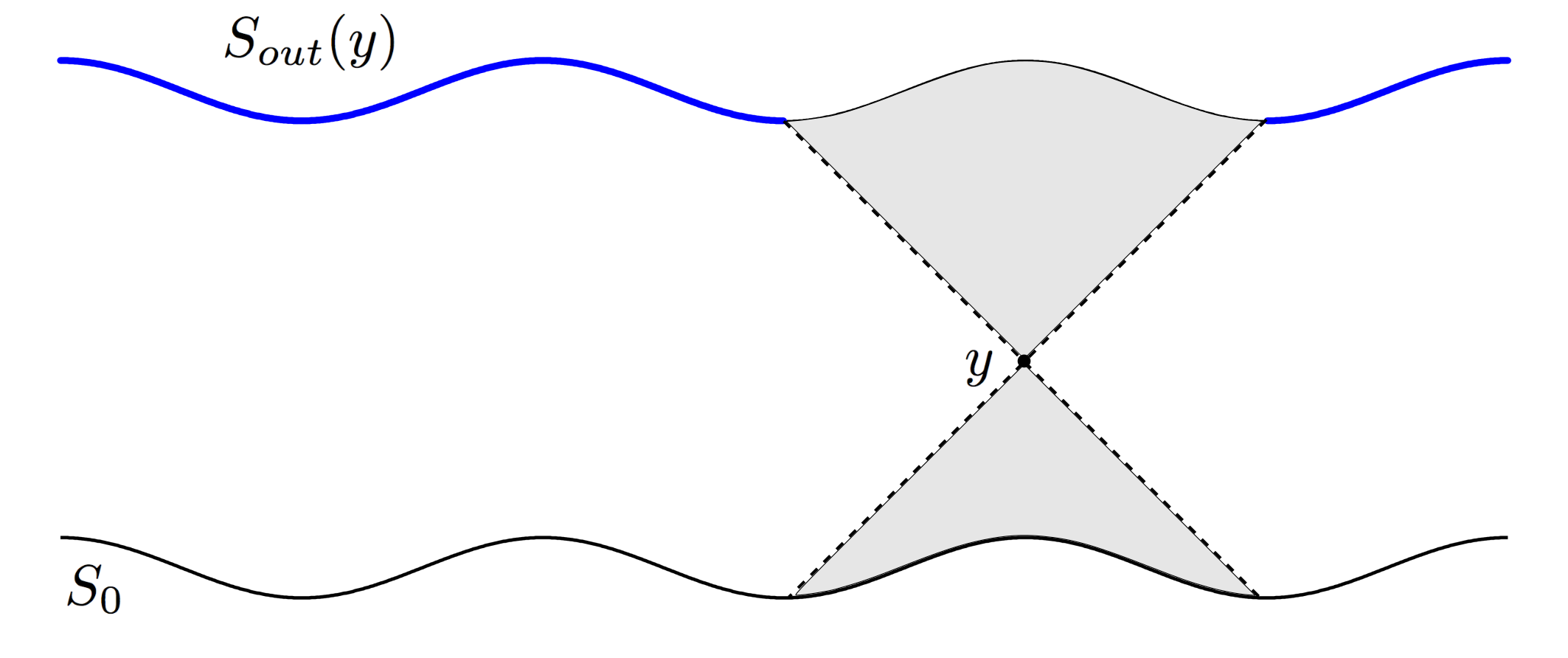}
\caption{ The spacelike hypersurface $S_{out}(y)$ in bold blue, for a given spacetime point $y$ between the initial hypersurface $S_0$ and final hypersurface $S$.}
\end{figure}

\iffalse
$$S_0$$
$$S_{out}(y)$$
$$y$$
\fi

Now consider an arbitrary spacetime point $y$ between $S_0$ and $S$. Kent constructs a number of hypersurfaces as a function of $y$, and the notation will deviate slightly from that used by Kent \cite{Kent20140241}, so to make the notation as clear as possible. First define the surface $\Lambda_S(y)$ as the set of points on the future light cone of the point $y$, up to and including the intersection of the light cone with the final surface $S$. Then define the surface $S_{out}(y)$ as the set of points $x\in S$ which lie \textit{outside} the future light cone of $Y$. Kent then calls the surface $S_\Lambda(y) := S_{out}(y)\cup\Lambda_S(y)$ the \textit{effective future boundary} of $y$. It includes all points on $S$ outside the future light cone of $y$ and the future light cone of $y$ itself up to the intersection with $S$.

To define probabilities, it is useful to consider $S_\Lambda(y)$ as the limit of a sequence of spacelike hypersurfaces, as follows. Consider a sequence of spacelike surfaces $S_i(y)$, with $i\in\mathbb{N}$, which contain the point $y$ and include all $x\in S_{out}(y)$ such that the distance from $x$ to any point $z\in S_{out}(y)\cap\Lambda_S(y)$ (where $S_{out}(y)$ meets the light cone) is greater than some distance $\epsilon_i$. If $\epsilon_i \to 0$ as $i \to \infty$ through some sequence, the condition that each $S_i(y)$ must remain spacelike ensures convergence to the unique surface $S_\Lambda(y) = \lim_{i\to\infty} S_i(y)$. A requirement of Kent's method is that the proposed solution must not be dependent on this limiting process.

Now for computing probabilities of intermediate spacetime measurements. As each $S_i(y)$ is a spacelike hypersurface in the future of $S_0$, one may use the Tomonaga-Schwinger formalism to evolve the initial state $\ket{\psi_0}$ on $S_0$ to some state $U_{S_iS_0}\ket{\psi_0}$ on the surface $S_i(y)$, ultimately being interested in the limit as $S_i(y) \to S_\Lambda(y)$. Now one needs to determine the probability of measuring a certain outcome $T_{\mu\nu}^j$ of the stress-energy tensor $T_{\mu\nu}$ at the point $y \in S_i(y)$, conditional on measuring $t_S(x)$ for all $x\in S_i(y) \cap S$ and conditional on the initial state $\ket{a}$. Notice that the post-selection takes place on $t_S(x)$ only for $x$ outside the light cone of $y$, implying that \textit{the conditional probabilities are blind to the part of the final boundary condition which falls within the timelike future of }$y$.

\begin{figure}[h]
\centering
\includegraphics[]{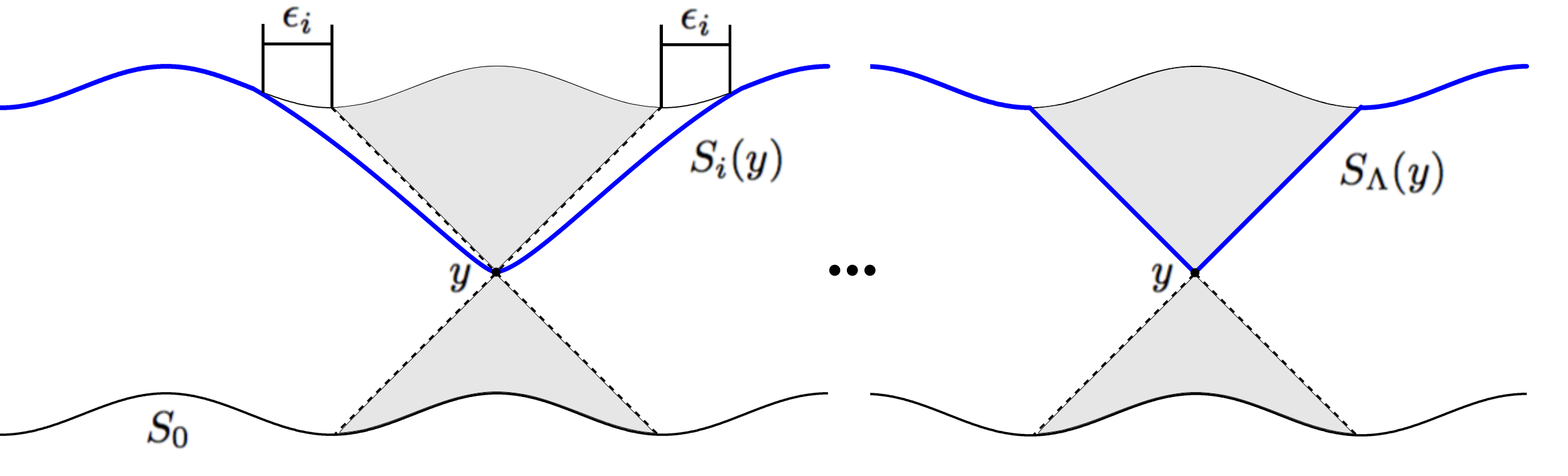}
\caption{Spacelike hypersurfaces $S_i(y)$ (left) converging, as $\epsilon_i\to 0$, to the effective future boundary $S_\Lambda(y)$ (right) for a given spacetime point $y$ between $S_0$ and $S$.}
\end{figure}

As demonstrated in section \ref{ABL}, this conditional probability can be decomposed, but with one difference. It cannot be said that $\textrm{Pr}\left(\,t_S(x) \bigm\vert T_{\mu\nu}^j \,, \ket{a} \right)=\textrm{Pr}\left(\,t_S(x) \,\, |\, T_{\mu\nu}^j \right)$. Previously, by L\"uders' rule, the intermediate measurement instantaneously collapses the state to a measurement eigenstate, so that the later probability of obtaining the final boundary condition depended only on the state after the intermediate measurement, not on the initial state. However, in the relativistic case, it is assumed that spacelike separated measurements commute, and that points on the final boundary condition which are spacelike separated from the point of measurement should not be influenced by the measurement. Thus, the initial state remains relevant at such points. With this pedantic correction, one may similarly formulate the conditional probability as in section \ref{ABL} equation (\ref{RelABL}) as
\begin{equation}
\textrm{Pr}\left(\,T_{\mu\nu}^j\bigm\vert t_S(x) , \ket{a}\right) = \frac{\textrm{Pr}\left(\,t_S(x) \bigm\vert T_{\mu\nu}^j \,, \ket{a} \right) \cdot \textrm{Pr}\left(\,T_{\mu\nu}^j\bigm\vert \ket{a}\right)}{\sum_{k}{\textrm{Pr}\left(\,t_S(x)\bigm\vert T_{\mu\nu}^k\,, \ket{a} \right)\cdot \textrm{Pr}\left(\,T_{\mu\nu}^k\bigm\vert \ket{a}\right)}}
\label{RelProbs}
\end{equation}
where $t_S(x)$ is considered for all $x \in S_i(y) \cap S$, and the sum over $j$ runs over all possible outcomes $T_{\mu\nu}^j$ of measuring the stress-energy. This procedure may be used to obtain conditional probabilities for measuring the stress-energy tensor for each point $y$ between $S_0$ and $S$. One may then construct the expectation value of the stress-energy, $\langle T_{\mu\nu}(y)\rangle$, in the usual way. That is, a probability has been associated to every possible outcome $T_{\mu\nu}^j$ of measuring the stress-energy tensor. By averaging over all possible $T_{\mu\nu}^j$ weighted by the corresponding conditional probability, the expectation value of $T_{\mu\nu}(y)$ is obtained as
\begin{equation}
 \langle T_{\mu\nu}(y)\rangle = \sum_j T_{\mu\nu}^j \cdot \textrm{Pr}\left(\,T_{\mu\nu}^j \bigm\vert t_S(x) , \ket{a}\right) 
\end{equation}
where the sum over $j$ again indexes over all possible measurement outcomes $T_{\mu\nu}^j$ of measuring the stress-energy.

In our universe defined between spacelike hypersurfaces $S_0$ and $S$, where $S$ is the limit of a convergent sequence of ever later spacelike hypersurfaces, Kent's framework thus constructs beable expectation values for the tensor field $\langle T_{\mu\nu}(y)\rangle$. The inferred physical reality of the system can then be succinctly described by
\begin{equation}
	\{ \langle T_{\mu\nu}(y)\rangle : S_0 < y < S\}
\end{equation}
where the notation implies that $y$ lies between the surfaces $S_0$ and $S$. 

\vspace{2mm}
\begin{center}
\textit{The Role of Photons in Kent's Relativistic Framework}
\end{center}
\vspace{2mm}
Our modified procedure for relativistic systems has a major implication which was, so far, not stated explicitly. Post-selecting only on the part of the final boundary condition that falls outside the light cone of a given spacetime point excludes a vast amount of information; in fact, by definition, \textit{only a particle moving at light speed could ever stay outside this light cone}: everything else in the universe which moves at less than light speed, such as massive particles, will eventually be swallowed up by the light cone as we let the final hypersurface $S$ tend to the infinite future.

Photons, or at least particles or field propagations which can keep pace with photons, thus play a vital role in Kent's solutions. They are the only particles which can ever escape the light cones of intermediate spacetime points as $S$ tends to future infinity, and thus are vital information carriers in Kent's framework. They are the only particles carrying information relevant for the conditional probabilities.

% SINGLE PHOTON TOY MODEL
\subsection{Toy Model: Single Photon Interaction}
\label{ToyModel1}

How exactly do photon-like particles knock the beables of the system into measurement eigenstates, despite the quantum state itself existing in superposition? Let us find out with a simple toy model presented by Kent \cite{Kent20140241}. Here, we will consider a massive system in a well-separated superposition of stationary, non-interacting states which are spatially well-separated in one dimension. The physical beable of the system will be mass density, that is, the physical reality of the system will ultimately be described by the expectation value of mass-density at every spacetime in the system.

To begin, one must address the behavior of photons themselves. It is true that photons may be most accurately described as having a non-negligible ``width" and corresponding wave packet, but for our purposes it is not unreasonable to model photons as purely point-like particles, perfectly localized in space, which always move at the speed of light. Furthermore, the toy models are presented in one spatial dimension, so that the photon wave function can be simply described as a delta function moving leftward or rightward at the speed of light. If the reader is put off by these assumptions, note that Kent has replaced these simplifications with the formalism of photon wave mechanics \cite{PhysRevA.96.062121} and achieved similar results.

This toy model considers an interaction between a slow-moving massive system in a superposition of two spatially well-separated states and a photon-like particle, which follows lightlike paths at the speed of light, taking $c=1$ in natural units. Denote the entire composite state as $\ket{\psi(t)}$ with subsystems $\ket{\psi_{S}(t)}$, the massive system in a superposition of states, and $\ket{\psi_\gamma(t)}$, the photon-like particle.

To construct the massive subsystem $\ket{\psi_{S}(t)}$ in a superposition, consider two states $\ket{\psi_1}$ and $\ket{\psi_2}$ each with mass $M$, which are thought of intuitively as ``mass clouds". For concreteness, let us take the wave functions $\psi_1(x)$ and $\psi_2(x)$ to be normalized Gaussian functions centered on position $x_1$ and $x_2$ with $x_1<x_2$, and of widths $\sigma_1$ and $\sigma_2$, respectively. The wave functions are required to be well separated, so that $\sigma_1,\sigma_2 \ll |x_2-x_1|$, implying that 
\begin{equation}
\bra{\psi_1}\ket{\psi_2}=\int \textrm{d}x \, \psi_1^{*}(x)\psi_2(x) \approx 0 \,.
\end{equation}
The details of the wave functions' spread about their center points are not central to the construction as long as the above condition holds; in particular, they need not be Gaussian. Assuming the above does hold, write the massive subsystem as the superposition $\ket{\psi_S}=a\ket{\psi_1}+b\ket{\psi_2}$ for $a,b\in \mathbb{C}$ with $|a|^2 + |b|^2=1$. For simplicity the dynamics and self-interactions of the massive subsystem are ignored and let the Hamiltonian $H_S$ of the subsystem be zero, so that $\ket{\psi_S(t)}$ has no dynamics when unperturbed by any other system.

Now consider the other part of the system, a single point-like particle $\ket{\psi_\gamma(t)}$ which moves at the speed of light, $c=1$, which is referred to as a ``photon". The initial wave function of the photon is modeled as
\begin{equation}
	\psi_\gamma(x;t)=\delta( x -(x_0 \pm t) )
\end{equation}
which, in words, is a delta function initially located at some point $x_0$, but which propagates to the right ($+t$) or to the left ($-t$) at the speed of light.

Interactions are simplified to ``bounces", as Kent puts it: upon an interaction of the photon and mass cloud 1 at position $x_1$, the photon reflects off the mass cloud and travels in the opposite direction,  and similarly for the other mass cloud. As Kent points out, this interaction indeed violates conservation of momentum, but if the mass clouds are considered to be macroscopic systems of non-negligible mass, the momentum imparted by the incoming photon will be negligible. Moreover, assume that interactions are certain to happen in the sense that a photon will always interact with one or the other of the mass clouds constituting the massive system.

Now consider a final boundary condition on $S$ in the form of a mass-energy distribution $T_S(x)=T_{\mu\nu}\hat n^\mu \hat n^\nu$ for all $x\in S$, where $T_{\mu\nu}$ is the stress-energy and $\hat n^\mu$ are the components of the unit vector normal to the tangent plane of $S$ at $x$. Kent postulates that the expectation value of mass-density $\rho(y)$ at any point in spacetime can be inferred by a single such stress-energy configuration $t_S(x)$ at late times.

For the initial conditions, let us consider the massive subsystem $\ket{\psi_S(t)}$ in a superposition of $\ket{\psi_1}$ and $\ket{\psi_2}$ as described above, with the photon propagating rightward from the far left of the entire system. Suppose that the photon will arrive at the point $x_1$ at some time $t_1$ and, if unreflected by the first mass cloud, will arrive at the point $x_2$ at time $t_2=t_1+(x_2-x_1)$. Using the position variable $x$ for the photon and $y$ for $\ket{\psi_S(t)}$, the wave function $\psi(x,y;t)$ of the system, for $t<t_1$, i.e. up until the time of the first possible ``bounce", is given by
\begin{equation}
	\psi(x,y;t)= \delta(x-(t+x_1-t_1))\cdot( a\psi_1(y) + b\psi_2(y)) \,.
\end{equation}

\begin{figure}[h]
\centering
\includegraphics[scale=0.5]{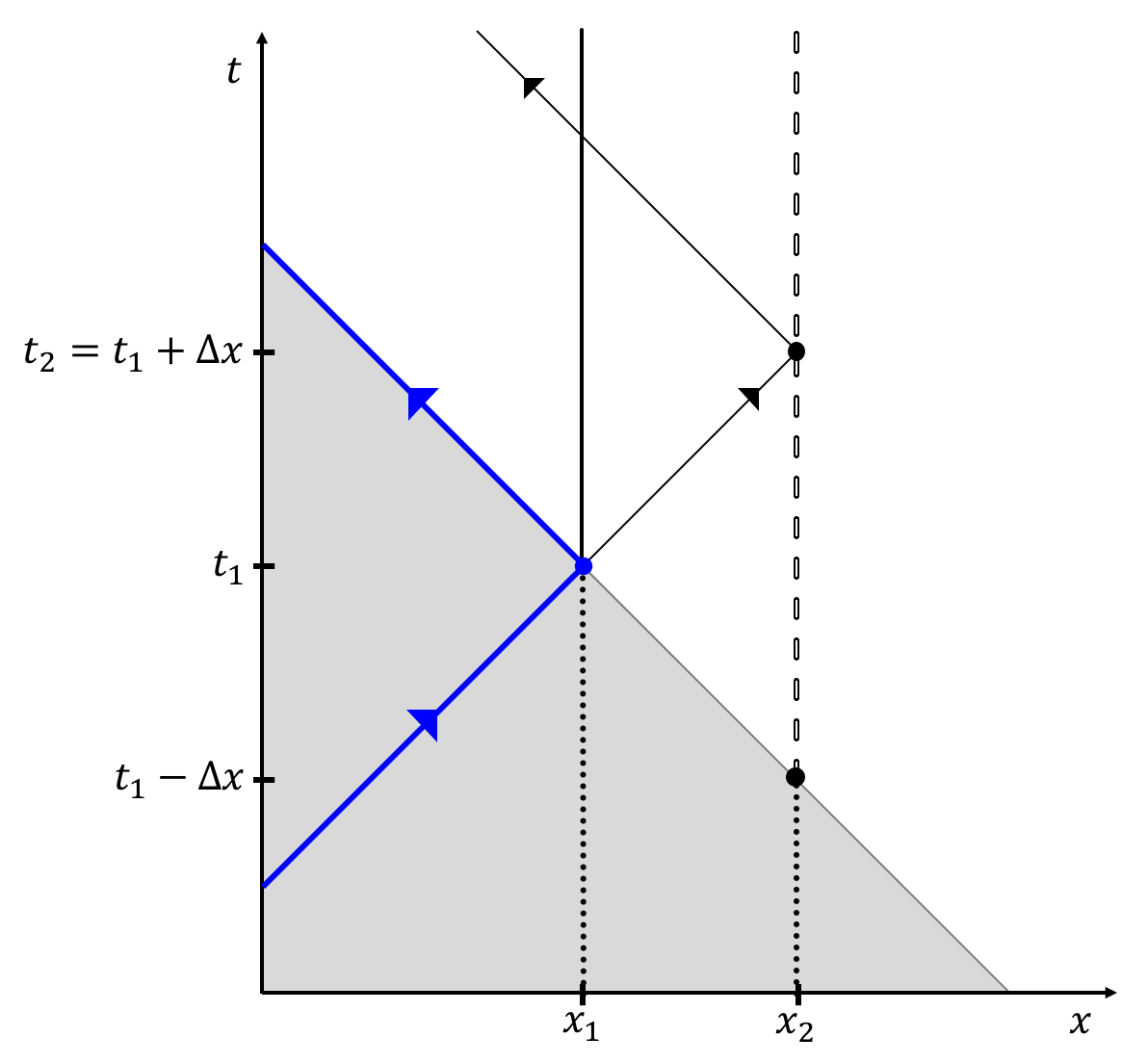}
\caption{ Single photon interaction. Diagonal lines represent possible photon paths, where the paths branch at $x_1$ and the wave function becomes a superposition. The thick blue line is the ``actual" path taken by the photon in the case where Nature randomly selects a final boundary condition with mass-energy distributed on the final hypersurface where it is intersected by the leftmost outgoing ray. The grey region depicts the ``region of indeterminacy", where mass density configurations represent Born rule weighted averages. The vertical lines at $x_1$ and $x_2$ show the temporal structure of the collapse of the mass density, and $\Delta x = x_2-x_1$. Dotted lines represent Born rule weighted proportions of mass density, solid vertical lines represent the complete presence of mass density, and dashed lines represent absence of mass density.}
\label{OnePhoton}
\end{figure}

At time $t_1$, the photon arrives at the first mass cloud at the point $x_1$. In our simplified interaction picture, one must account for the two possibilities corresponding to the superposition between the two mass clouds. In the first scenario, the photon interacts with the first mass cloud at $x_1$, reversing its direction as a result. In the second scenario, the photon interacts at time $t_2$ with the second mass cloud after propagating to $x_2$. For the intermediate times between $t_1<t<t_2$, the wave function of the system thus takes the form
\begin{equation}
	\psi(x,y;t)= \delta(x-(t_1+x_1-t))\cdot a\psi_1(y)  + \delta(x-(t+x_1-t_1))\cdot b\psi_2(y) \, .
\end{equation}
The situation is depicted in figure \ref{OnePhoton} assuming that in fact the photon reflects off of the first mass cloud rather than the second one. Finally, after time $t_2$ the photon heading to the second mass cloud will have also interacted and reversed its direction, implying the wave function of the system for times $t>t_2$ as
 \begin{equation}
	\psi(x,y;t)= \delta(x-(t_1+x_1-t))\cdot a\psi_1(y)  + \delta(x-(t_2+x_2-t))\cdot b\psi_2(y).
\end{equation}

Let us consider a universe in which Nature selects a certain final boundary condition $t_S(x)$ for the late-time mass-energy configuration which indicates that the photon indeed reflects off the first mass cloud - that is, one finds localized mass-energy density along the leftmost outgoing light ray $\delta(x-(t_1+x_1-t))$ at late times in $t_S(x)$. The goal is now to define expectation values of mass density $\langle \rho(y)\rangle$ for all intermediate points in spacetime, with particular interest in the mass density at the spatial points $x_1$ and $x_2$ corresponding to the centers of the mass clouds.

Postulating that the photon moves along lightlike paths greatly simplifies the conditional probabilities of measuring any given value of mass density at intermediate times, since in this model there are only two possible outcomes of a fictional late-time measurement of mass-energy: that in which the photon mass-energy is found in the position which corresponds to having reflected from the first or second mass cloud. Moreover, the leftmost outgoing light ray $\delta(x-(t_1+x_1-t))$ encodes both possible outcomes - if mass-energy is found along this light ray at late time, it must be that the photon reflected from the first mass cloud; if no mass-energy is found along the light ray, the photon must have reflected from the second mass cloud. For an arbitrary point $y$ in spacetime, if the leftmost outgoing light ray is present on the surface $S_{out}(y)$ (outside the light cone of $y$), then the presence or absence of mass-energy along that light ray makes definite whether or not the photon ``actually" reflected from the first mass cloud. Thus, if the leftmost outgoing light ray falls outside the future light cone of a given intermediate point $y$, the mass density will be forced into taking a definite value localized at either $x_1$ or $x_2$, despite the wave function of the massive subsystem persisting in superposition. In this example, where Nature randomly selected the final mass-energy configuration with mass-energy in the support of the leftmost outgoing light ray, the mass density must be found localized around $x_1$ for such points $y$.

Here it is useful to go beyond Kent's exposition and introduce the concept of a \textit{region of indeterminacy}, depicted as the grey region in figure \ref{OnePhoton}. In the context of a system in a superposition state, the region of indeterminacy is the set $\{y\}$ of points in spacetime for which the surface $S_{out}(y)$ does \textit{not} contain the relevant information which describes the final boundary condition for the massive subsystem. Recalling our discussion from section \ref{BackToBorn}, for all spacetime points in the region of indeterminacy, every possible final boundary condition appears the same - all of the relevant information is contained within the forward light cone, inaccessible in the conditional probability. It was shown that in this case, the conditional probability defaults to the Born rule given the unitarily evolved initial state. 

In this case, the region of indeterminacy can be explicitly stated as the set of spacetime points
\begin{equation}
\{y'=(t',x') :  t' < t_1-(x'-x_1)  \}.
\end{equation}
For spacetime points in this region of indeterminacy, the mass-density beables remain in a configuration corresponding to their Born rule weighted average, with $|a|^2M$ proportion of the mass distributed around $x_1$ and $|b|^2$ proportion distributed around $x_2$. For spacetime points outside the region of indeterminacy, in a universe in which Nature selects a final boundary condition with mass-energy on the leftmost outgoing ray, the mass is found completely localized around $x_1$. 

Thus, the mass-density beables $\rho(y)$ describing the physical reality of the system are summarized in figure \ref{OnePhoton} and summarized as follows. For spacetime points $y'=(t',x')$ falling within the region of indeterminacy, the mass density distribution takes the form
\begin{equation}
\langle \rho(x';t') \rangle = M |\alpha|^2 |\psi_1(x')|^2 + M |\beta|^2 |\psi_2(x')|^2
\end{equation}
while for spacetime points $y=(t,x)$ outside the region of indeterminacy, the mass density distribution takes the form
\begin{equation}
\langle \rho(x';t') \rangle = M |\psi_1(x')|^2 .
\end{equation}
Similarly, if Nature had chosen a final boundary condition with mass-energy in the support of the rightmost outgoing light ray, the mass density distribution would be distributed like $|\psi_2(x')|^2$ around $x_2$ for spacetime points outside the region of indeterminacy.

An interesting observation about this toy model is that the rightmost outgoing light ray corresponding to the photon interacting with the second mass cloud at $x_2$ is never relevant, the presence or absence of the photon along the the first outgoing light ray is sufficient to determine the conditional probabilities for intermediate measurements of mass density.

As depicted in figure \ref{OnePhoton}, ``news" of the interaction at $x_1$ and the ensuing collapse of mass density does not propagate instantaneously across the universe at time $t_1$ (the time of interaction) as one might intuitively expect. In fact, one sees a \textit{collapse defined by the leftmost outgoing light ray}, so that points to the right of $x_1$ in collapse before $t_1$ and points to the left of $x_1$ collapse later in time than $t_1$.

There is another, seemingly awkward, conclusion---during the time interval $t_1-\Delta x < t < t_1$ it seems that the mass density is only ``partly present" as Kent puts it \cite{Kent20140241}. That is, for a time-slice of the universe in this time interval, the mass density does not seem to be fully realized, with only $|\alpha|^2<1$ percentage of the total mass present. This indeed appears awkward, though one must recall that, in a sense, a ``time-slice" of the universe is not a relativistic concept. Even if the collapse did occur instantaneously in one reference frame, it necessarily would not appear so in any other boosted frame of reference.

%% TWO PHOTON MODEL
\subsection{Toy Model: Two Photon Interaction}
\label{ToyModel2}

Intuitions may be developed further with a small addition to the first toy model of section \ref{ToyModel1}, in which a second photon is introduced to the system. Ultimately, a similar story will emerge: but with a corresponding collapse of mass density across a light cone structure instead of across a single light ray, and with a corresponding triangular region of indeterminacy.

As before, consider the state $\ket{\psi_S(t)}$ initially in a superposition of two spatially well-separated states $\ket{\psi_1}$ localized around position $x_1$ and $\ket{\psi_2}$ around $x_2$, such that, initially, $\ket{\psi_S(t)}= a\ket{\psi_1} + b\ket{\psi_2}$ for $a,b\in \mathbb{C}$ with $|a|^2+|b|^2=1$. Each state again has mass $M$. The dynamics and self-interactions of $\ket{\psi_S(t)}$ are ignored so that the state doesn't evolve in time when unperturbed. As before, interactions between $\ket{\psi_S(t)}$ and incoming photons are modeled by a simple bounce of the photon off the system. 

\begin{figure}[h]
\centering
    \includegraphics[scale=0.5]{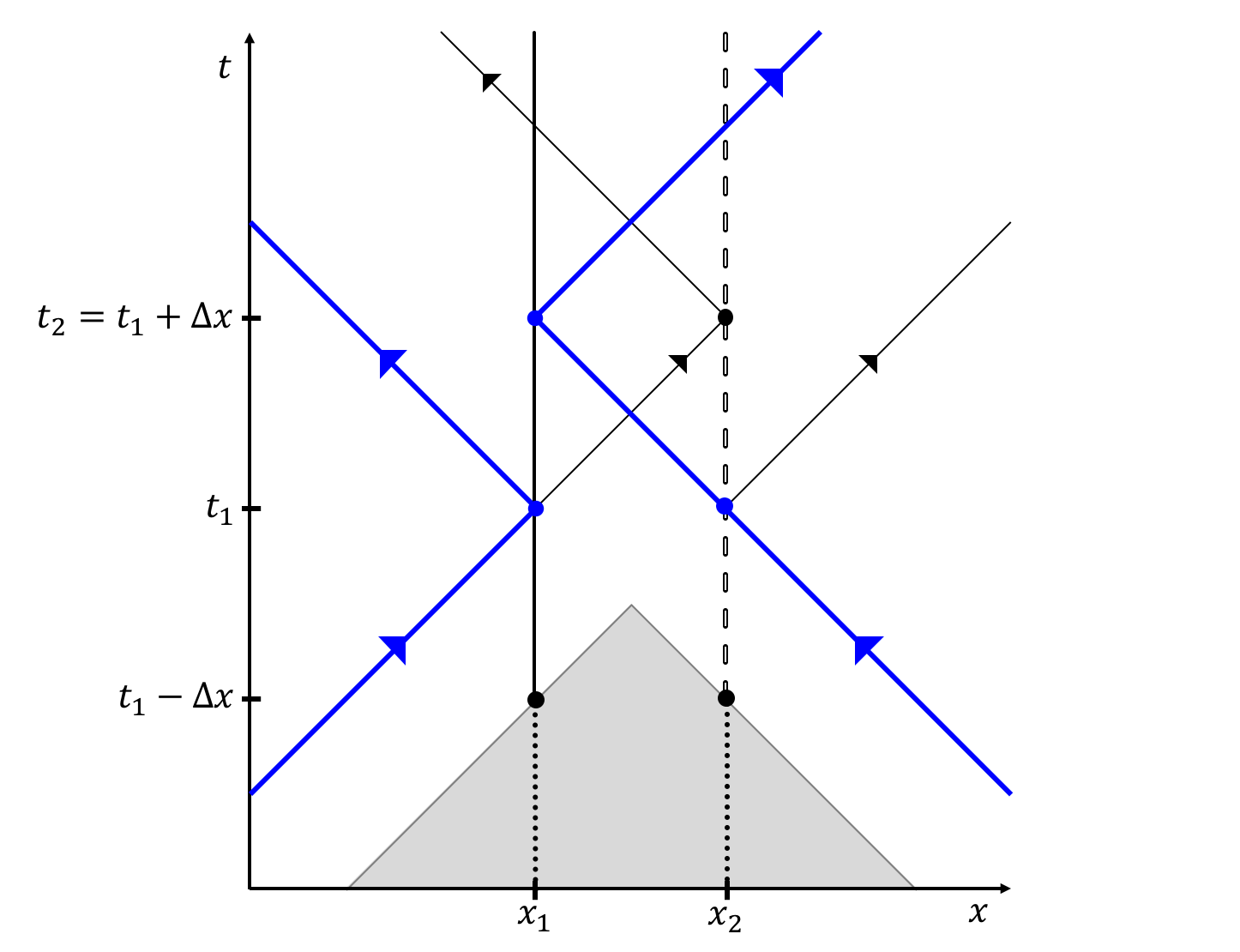}
    \caption{Two photon interaction. Diagonal lines represent possible photon paths. The thick blue line is the ``actual" paths taken by the photons, one of the two possible outcomes which was chosen at random by Nature. The grey region depicts the ``region of indeterminacy", triangular this time, where mass density configurations represent Born rule weighted averages. The vertical lines at $x_1$ and $x_2$ show the temporal structure of the collapse of the mass density, and $\Delta x = x_2-x_1$. Dotted lines represent Born rule weighted proportions of mass density, solid vertical lines represent the complete presence of mass density, and dashed lines represent absence of mass density.}
\label{TwoPhotons}
\end{figure}

Now suppose there are two photons at play in the system, the states of which are denoted as $\ket{\psi_{\gamma1}(t)}$ and $\ket{\psi_{\gamma2}(t)}$. The first photon $\ket{\psi_{\gamma1}(t)}$ approaches the system from the left, reaching the point $x_1$ at time $t_1$ as before, and now suppose that the second photon $\ket{\psi_{\gamma2}(t)}$ approaches the system from the right, reaching the point $x_2$ at the same time $t_1$. Allowing $x$ to label the position variable for $\ket{\psi_{\gamma1}}$, $y$ for $\ket{\psi_S}$, and now also $z$ for $\ket{\psi_{\gamma2}}$, the wave function for the system for times $t<t_1$ becomes
\begin{equation}
	\psi(x,y,z;t) = \delta(x-(x_1+t-t_1))\cdot\delta(z-(x_2-t+t_1))\cdot(a\psi_1(y)+b\psi_2(y)).
\end{equation}

At time $t_1$, the rightward traveling photon $\ket{\psi_{\gamma1}}$ arrives at mass cloud one, while the leftward traveling photon $\ket{\psi_{\gamma2}}$ simultaneously arrives at mass cloud two. As $\ket{\psi_S}$ is in a superposition, one must again account for the two possibilities in which the superposed state is ``found by a photon" around one or the other mass cloud. In either case, letting $X=1,2$, one of the photons will interact with mass cloud $X$ at time $t_1$ and reverse its direction, while the other continues propagating toward mass cloud $X$ up until interaction with mass cloud $X$ at time $t_2=t_1+(x_2-x_1)$. Thus, for $t_1<t<t_2$ the wave function of the system takes the form below
\begin{equation}
\begin{split}
	\psi(x,y,z;t) &=\delta(x-(x_1-t+t_1))\cdot\delta(z-(x_2-t+t_1))\cdot a\psi_1(y)\\
	&\,+\delta(x-(x_1+t-t_1))\cdot\delta(z-(x_2+t-t_1))\cdot b\psi_2(y) \,.\\
\end{split}
\end{equation}
Finally, upon reaching the mass cloud $X$ at time $t_2$, the other photon interacts and reverses its direction as well. Then for $t>t_2$, the wave function of the system assumes the form
\begin{equation}
\begin{split}
	\psi(x,y,z;t) &=\delta(x-(x_1-t+t_1))\cdot\delta(z-(x_1+t-t_2))\cdot a\psi_1(y)\\
	&\,+\delta(x-(x_2-t+t_2))\cdot\delta(z-(x_2+t-t_1))\cdot b\psi_2(y) \,.\\
\end{split}
\end{equation}

Now, for the final boundary condition, suppose that Nature selects an asymptotic late-time mass-energy distribution which indicates that the photons $\ket{\psi_{\gamma1}}$ and $\ket{\psi_{\gamma2}}$ reflected from mass cloud one located at $x=x_1$, though of course we could analogously consider the photons reflecting from mass cloud two.

The conditional probabilities are again very simple - it remains true that the leftmost outgoing light ray corresponding to $\ket{\psi_{\gamma1}}$ reflecting off mass cloud one encodes all the relevant information about the system. Also similar to the first toy model: the other light ray for $\ket{\psi_{\gamma1}}$ which reflects from mass cloud two is never needed, as it lags behind the leftmost light ray and is thus ``harder to see" in the conditional probability.  However, there is now another set of two light rays from $\ket{\psi_{\gamma2}}$. The rightmost outgoing ray corresponding to $\ket{\psi_{\gamma2}}$ reflecting from mass cloud two also encodes all of the relevant information about the system in the sense that if mass-energy is found along this light ray, the mass density of the massive subsystem must be found centered around mass cloud two, and otherwise, it must be found centered around mass cloud 1. Thus, one now has a collapse along the rightmost outgoing light ray \textit{as well as} the leftmost outgoing light ray. Any point in spacetime $y$ which falls in the future-time side of at least one of these light rays will ``see" the relevant information in the late-time mass-energy distribution in the surface $S_{out}(y)$ used in the conditional probability. Thus, the region of spacetime which does \textit{not} see the collapse must lie in the past-time side of both collapses across the leftmost and rightmost outgoing light rays, resulting in a triangular region of indeterminacy.

In this case, the region of indeterminacy is formulated as the set of spacetime points $y'$ defined by
\begin{equation}
\label{ROI2}
\{ y'=(t',x') : t' < t_1 + (x'-x_2) \textrm{ and } t' < t_1- (x'-x_1) \}.
\end{equation}

Thus, the mass-density beables $\rho(y)$ describing the physical reality of the system are summarized in figure \ref{TwoPhotons} and summarized as follows. For spacetime points $y'=(t',x')$ falling within the region of indeterminacy (\ref{ROI2}), the mass density distribution takes the form
\begin{equation}
\langle \rho(x';t') \rangle = M |\alpha|^2 |\psi_1(x')|^2 + M |\beta|^2 |\psi_2(x')|^2
\end{equation}
while for spacetime points $y=(t,x)$ outside the region of indeterminacy, the mass density distribution takes the form
\begin{equation}
\langle \rho(x';t') \rangle = M |\psi_1(x')|^2 .
\end{equation}
Similarly, if Nature had chosen a final boundary condition with mass-energy in the support of the light rays reflecting from the mass cloud two instead of mass cloud one, the mass density distribution would be distributed like $|\psi_2(x')|^2$ around $x_2$ for spacetime points outside the region of indeterminacy.

As made evident in figure \ref{TwoPhotons}, in this case one gets a full collapse of mass density at time $t_1-\Delta x$. This is due to the two photons arriving at their respective system precisely at time $t_1$ in the rest frame of superposed massive system. If one of the photons were to arrive before or after the other, the corresponding region of indeterminacy would be shifted so that the mass density collapses at $x_1$ and $x_2$ would no longer occur at the same time in the rest frame. Thus, the ``missing mass" phenomenon of the first model remains present.

Introducing more photons into the system thus has the effect of shrinking the region of indeterminacy, or equivalently making the mass density beables describing physical reality localized at more points in spacetime as opposed to being distributed between $x_1$ and $x_2$. In most other regards, the picture is very similar to the single photon model. Both models have in common the feature that the actual mass density of the system remains distributed between the states in superposition until interaction with an ``information carrying" photon, at which point they collapse into localized configurations.

% CONCLUSION
\newpage
\section{Conclusion}
Adrian Kent has proposed a mathematically precise interpretation of quantum theory which solves the quantum reality problem. By considering a sample space of physical beable distributions in the asymptotic infinite future, with a well defined probability distribution over those distributions, one can postulate that the physical reality of the quantum system corresponds to one such distribution chosen at random from the sample space. Entire histories of the beables' actual expectation values are then inferred from the final distribution, making definite the physical reality of the quantum system at all points in spacetime. Moreover, his framework defines these histories using fully Lorentz covariant rules, ultimately producing Lorentz covariant descriptions of quantum reality. 

In this paper, Kent's framework was presented and worked out in detail for a number of systems. As increasingly sophisticated models are considered \cite{Kent20140241,PhysRevA.96.062121}, Kent's interpretation continues to provide a coherent and illuminating description of physical reality in quantum systems; of course, with no rigorous formulation of relativistic quantum theory known at present, a thorough analysis of Kent's interpretation must ultimately wait. Nonetheless, his interpretation of quantum theory endows quantum reality with physical existence and mathematical precision, solving the quantum reality problem, and providing us all with an enlightening perspective of the quantum world.

\subsection*{Acknowledgements}
I sincerely thank Jeremy Butterfield for his supervision, editing, and his Michaelmas 2017 term class ``Philosophical Aspects of Quantum Field Theory" during Part III of the Mathematical Tripos at the University of Cambridge. Comments passed on from Adrian Kent were much appreciated and greatly improved the text. Gijs Leegwater was very kind in sharing his time to explain non-locality in Kent's solutions. Finally, I thank Sam Crawford for his talk on interpretations of quantum theory and Matthew Horner for stimulating discussions. This material is based upon work supported by the National Science Foundation Graduate Research Fellowship under Grant No. DGE - 1656518. 

\bibliography{biblio}{}
\bibliographystyle{ieeetr}

\end{document}